\documentclass{aa}  

\usepackage{graphicx}
\usepackage{txfonts}
\usepackage{threeparttable}
\usepackage[normalem]{ulem}
\usepackage[version=4]{mhchem} 
\usepackage{float}
\usepackage{placeins}
\usepackage{capt-of}
\providecommand{\as}{^{\prime\prime}}
\DeclareUnicodeCharacter{2212}{-}
\usepackage{hyperref}
\hypersetup{pageanchor=false}
\usepackage[export]{adjustbox}
\makeatletter
\renewcommand*\aa@pageof{, page \thepage{} of \pageref*{LastPage}}
\makeatother

\begin{document} 

   \title{ALMA high-resolution observations unveil planet formation shaping molecular emission in the PDS~70 disk}
   \authorrunning{L. Rampinelli et al.}
   \titlerunning{ALMA high-resolution line emission observations of the PDS~70 disk}
   \author{L. Rampinelli \inst{1}, S. Facchini \inst{1}, M. Leemker\inst{1}, J. Bae \inst{2}, M. Benisty \inst{3,4}, R. Teague\inst{5}, C. J. Law\inst{6,7}, K. I. \"Oberg\inst{8}, B. Portilla-Revelo\inst{9, 10}, and A. J. Cridland\inst{11}
   }

   \institute{Dipartimento di Fisica, Universit\`a degli Studi di               Milano, Via Celoria 16, 20133 Milano, Italy\\
            \email{luna.rampinelli@unimi.it}
        \and 
            Department of Astronomy, University of Florida, Gainesville, FL 32611, USA
        \and
            Université Côte d’Azur, Observatoire de la Côte d’Azur, CNRS, Laboratoire Lagrange, F-06304 Nice, France
        \and
            Université Grenoble Alpes, CNRS, IPAG, 38000 Grenoble, France
        \and 
            Department of Earth, Atmospheric, and Planetary Sciences, Massachusetts Institute of Technology, Cambridge, MA 02139, USA
        \and
            NASA Hubble Fellowship Program Sagan Fellow
        \and
            Department of Astronomy, University of Virginia, Charlottesville, VA 22904, USA
        \and
            Center for Astrophysics | Harvard \& Smithsonian, 60 Garden St., Cambridge, MA 02138, USA
        \and 
            Kapteyn Astronomical Institute, University of Groningen, PO Box 800, 9700 AV Groningen, The Netherlands
        \and
            Department of Astronomy \& Astrophysics, The Pennsylvania State University, 525 Davey Laboratory, University Park, PA 16802, USA
        \and
            Universit\"ats-Sternwarte M\"unchen, Ludwig-Maximilians-Universit\"at, Scheinerstr. 1, 81679 M\"unchen, Germany
            }

   \date{Received 22 February 2024 ; accepted 2 July 2024}

 
  \abstract{
    With two directly detected protoplanets, the PDS~70 system is a unique source in which to study the complex interplay between forming planets and their natal environment. The large dust cavity carved by the two giant planets can affect the disk chemistry, and therefore the molecular emission morphology. On the other hand, chemical properties of the gas component of the disk are expected to leave an imprint on the planetary atmospheres. In this work, we reconstruct the emission morphology of a rich inventory of molecular tracers in the PDS~70 disk, and we look for possible chemical signatures of the two actively accreting protoplanets, PDS~70b and c. We leverage Atacama Large Millimeter/submillimeter Array (ALMA) band 6 high-angular-resolution and deep-sensitivity line emission observations, together with image and $uv$-plane techniques, to boost the detection of faint lines. We robustly detect ring-shaped emission from \ce{^12CO}, \ce{^13CO}, \ce{C^18O}, \ce{H^13CN}, \ce{HC^15N}, DCN, \ce{H2CO}, CS, \ce{C2H}, and \ce{H^13CO^+} lines in unprecedented detail. Most of the molecular tracers show a peak of the emission inside the millimeter dust peak. We interpret this as the direct impact of the effective irradiation of the cavity wall, as a result of the planet formation process. Moreover, we have found evidence of an O-poor gas reservoir in the outer disk, which is supported by the observations of bright C-rich molecules, the non-detection of SO, and a lower limit on the $\mathrm{CS/SO}$ ratio of $\sim1$.  Eventually, we provide the first detection of the \ce{c-C3H2} transitions at 218.73~GHz, and the marginal detection of an azimuthal asymmetry in the higher-energy \ce{H2CO} (3$_{2,1}$-2$_{2,0}$) line, which could be due to accretion heating near PDS~70b.}
    

   \maketitle
%
\section{Introduction} \label{sec:intro}

In the past decade, advances in exoplanet detection and characterization have allowed for demographic insights. They have revealed a broad diversity in exoplanet properties, and thus imply that the outcomes of planet formation are similarly diverse \citep[][and references therein]{suzuki2016exoplanet, thompson2018planetary, fulton2021california, vigan2021sphere, currie2022direct, lissauer2023exoplanet}.
Mature planetary systems store important markings of their formation history: for example, the composition of their atmosphere may be crucial in determining the chemical inheritance from their natal environment; that is, the protoplanetary disk \citep[][and references therein]{oberg2021astrochemistry, nomura2022isotopic, oberg2023protoplanetary}. In recent years, detailed information about the chemical properties and composition of gas giant exoplanets has been inferred from transmission and thermal emission spectra thanks to powerful ground-based facilities, or space missions \citep[][]{greene2016characterizing, hinkley2022jwst, currie2022direct, guillot2022giant}. Analyzing the obtained spectra with radiative transfer thermo-chemical models has revealed unprecedented details about planetary atmospheres, constraining fundamental properties such as the C/O elemental ratio \citep{molliere2015model} or carbon fractionation \citep{zhang202113co}. 

On the one hand, planet formation is strongly influenced by the chemical and physical processing of the disk material during the disk evolution, which is expected to affect the location, timescale, mass budget, or chemical composition of forming planets. On the other hand, a forming planet is also expected to leave an imprint on the disk, such as substructures in the dust and gas distributions, chemical signatures, and kinematics perturbations, which have been observed and used to predict the planet's properties \citep[][and references therein]{bae2022structured, benisty2022optical, pinte2022kinematic}.

In this context, high spatial and spectral resolution observations of molecular lines with ALMA can be leveraged to detect both kinematic and chemical footprints of planet formation. The kinematic variations induced onto the Keplerian rotation pattern are observed as localized spatial and spectral features \citep{perez2015planet, pinte2018kinematic, Casassus2019kinematic, pinte2019kinematic, teague2019meridional, izquierdo2022new, pinte2022kinematic, izquierdo2023maps, stadler2023kinematically}. Chemical signatures induced by planetary accretion shocks can be targeted through specific molecular lines \citep{cleeves2015indirect, law2023so}, or chemical footprints of planet formation can be observed through substructures into the gas distribution \citep{oberg2021molecules, bib:facchini2021chemical, nomura2021high, law2021molecules, bae2022cpd, jiang2023chemical}. 

Planetary signatures in molecular line emission observations complement constraints on planet formation coming from studies on the dust distribution, which have been largely explored in the past few years \citep{andrews2018disk, long2018gaps, andrews2020observations, cieza2021ophiuchus, sierra2021molecules, bae2022structured}, resulting in the recent detection of a circumplanetary disk in the PDS~70 system through ALMA continuum observations \citep{benisty2021circumplanetary}. Moreover, molecular line emission observations can provide a way to access open questions related to the astrochemical characterization of planet-forming disks, such as elemental composition \citep{bosman2021molecules, cataldi2021molecules}, evolution of organic chemistry \citep{guzman2021molecules, ilee2021molecules}, and in particular sulfur chemistry \citep{le2021molecules}, or snowline location.

In this context, the PDS~70 system is the best candidate to further explore the impact of planet formation on the chemical structure of the disk, and on the other hand how the chemical properties of the disk can result in the final atmospheric composition of the accreting planets \citep{cridland2023planet}, as it is the first source for which a multiwavelength direct detection of two forming planets has been presented, in near-infrared \citep{keppler2019highly, christiaens2019separating, mesa2019vlt}, H$\alpha$ line \citep{wagner2018magellan, haffert2019two}, ultraviolet \citep{zhou2021hubble}, and millimeter and submillimeter observations \citep{isella2019detection, benisty2021circumplanetary}. PDS~70 is a $\sim$5 Myr old T Tauri star with a 0.87~M$_\odot$ mass \citep{paxton2010modules, hashimoto2015structure, choi2016mesa, keppler2018discovery, long2018differences, keppler2019highly} at a distance of $\sim$113 pc \citep{brown2018gaia} in the Upper Centaurus-Lupus subgroup. \cite{bib:facchini2021chemical} presented the first chemical inventory of the PDS~70 disk, showing a complexity of morphologies and emitting properties, and evidence suggesting a gas-phase $\mathrm{C}/\mathrm{O}>1$. \cite{law2024mapping} extracted the high-resolution vertical gas structure of the PDS~70 disk, showing strong evidence of a prominent cavity wall at the \ce{^12CO}-emitting height, which could directly impact the molecular emission morphology. Moreover, \cite{Perotti2023water} found water in the JWST spectrum of the inner disk of PDS~70, which could affect potential terrestrial planets being formed.

In this work, we analyze ALMA band~6 line emission observations of the planet-hosting disk around PDS~70. We present new high-angular-resolution observations of multiple molecular lines, providing an overview of the molecular complexity in the PDS~70 disk in unprecedented detail. In particular, we suggest that the effect of the strongly irradiated cavity wall is playing an important role in shaping the molecular emission. 
In Sect.~\ref{sec:data}, we describe the observations together with calibration and imaging procedures.
In Sect.~\ref{sec:boost}, we describe the image-plane and $uv$-plane techniques we applied to boost faint lines. 
In Sect.~\ref{sec:results_molecules}, we outline the analyses we performed in order to build a chemical overview of the observed molecules. In Sect.~\ref{sec:discussion}, we interpret the observed emission morphologies of the analyzed molecular tracers in light of the ongoing process of planet formation.
We summarize the results and discussion in Sect.~\ref{sec:conclusions}.

\section{Observations and data reduction} \label{sec:data}

\begin{table*}
  \renewcommand{\arraystretch}{1.2}
  \centering
  \caption{
  Listed imaged lines, corresponding rest frequencies, imaging parameters (channel width, beam, RMS), and line fluxes.}
  \label{tab:imaging_molecules}
  \begin{threeparttable}
  \begin{tabular}{ c c c c c c c c c}
      \hline
      \hline
      Species & Transition\tnote{a} & Frequency &  $\Delta v$ & Beam & RMS & Flux\tnote{b} & Velocity range\tnote{c} & $\epsilon$ factor\\
      &  & [GHz] & [km~s$^{-1}$] &  & [mJy~beam$^{-1}$] & [mJy km~s$^{-1}$] &  [km~s$^{-1}$] & \\
      \hline
      $^{12}$CO & 2-1 & 230.5380 & 0.1 & $0\farcs28 \times 0\farcs22$ & 0.54 & 6175 $\pm$ 23 & 0.0-10.3 & 0.30\\
      $^{13}$CO & 2-1 & 220.3986 & 0.2 & $0\farcs30 \times 0\farcs24$ & 0.42 & 1859 $\pm$ 12 &  1.7-9.1 & 0.32\\
      C$^{18}$O & 2-1 & 219.5603 & 0.2 & $0\farcs31 \times 0.24$ & 0.33 & 466 $\pm$ 7 & 2.1-8.5 & 0.31 \\
      H$_2$CO & 3$_{0,3}$-2$_{0,2}$ & 218.2221 & 0.4 & $0\farcs29 \times 0\farcs23$ & 0.19 & 708 $\pm$ 7 &  1.9-9.1 & 0.30\\
      H$_2$CO & 3$_{2,1}$-2$_{2,0}$ & 218.7600 & 0.4 & $0\farcs29 \times 0\farcs23$ & 0.20 & 57 $\pm$ 6 & 2.7-8.3 & 0.30\\
      C$_2$H & $3_{7/2, 4}$-$2_{5/2, 3}$ & 262.0042 & 0.4 & $0\farcs20 \times 0\farcs15$ & 0.16 & 723 $\pm$ 10 & 0-9.1\tnote{d} & 0.23\\
      C$_2$H & $3_{7/2, 3}$-$2_{5/2, 2}$ & 262.0064 & 0.4 & $0\farcs20 \times 0\farcs15$ & 0.16 & 577 $\pm$ 10 & 0-9.1\tnote{d} & 0.23\\
      H$^{13}$CN & 3-2 & 259.0117 & 0.4 & $0\farcs20 \times 0\farcs15$ & 0.13 & 360 $\pm$ 8 & 1.9-9.1 & 0.22\\
      HC$^{15}$N & 3-2 & 258.1571 & 0.4 & $0\farcs20 \times 0\farcs15$ & 0.15 & 176 $\pm$ 7 & 2.7-8.3 & 0.22\\
      DCN & 3-2 & 217.2386 & 0.4 & $0\farcs29 \times 0\farcs23$ & 0.22 & 230 $\pm$ 6 & 2.3-8.7 & 0.30\\
      H$^{13}$CO$^+$ & 3-2 & 260.2553 & 0.4 & $0\farcs21 \times 0\farcs15$ & 0.16 & 372 $\pm$ 7 & 2.3-8.7 & 0.23\\
      CS & 5-4 & 244.9355 & 1.4 & $0\farcs22 \times 0\farcs15$ & 0.06 & 445 $\pm$ 6 & 2.7-8.3 & 0.23\\
      \ce{c-C3H2} & 3$_{2,1}$-2$_{1,2}$ & 244.2221 & 1.4 & $0\farcs22 \times 0\farcs16$ & 0.07 & 16 $\pm$ 2 & 1.3-9.7 & 0.52\\
      \ce{c-C3H2} & 7$_{1,6}$-7$_{0,7}$ & 218.7327 & 0.4 & $0\farcs29 \times 0\farcs23$ & 0.20 & 10 $\pm$ 3\tnote{e} &  1.5-8.3 & 0.52 \\
      \ce{c-C3H2} & 7$_{2,6}$-7$_{1,7}$ & 218.7327 & 0.4 & $0\farcs29 \times 0\farcs23$ & 0.69 & 10 $\pm$ 3\tnote{e} &  1.5-8.3 & 0.52 \\
      SO & 6$_7$-5$_6$ & 261.8436 & 0.2 & $0\farcs20 \times 0\farcs15$ & 0.91 & $<$ 149\tnote{f} & 1.9-9.1\tnote{g}&--\\
      SO & 6$_6$-5$_5$ & 258.2558 & 0.5 & $0\farcs21 \times 0\farcs15$ & 0.66 & $<$ 106\tnote{f} & 2.0-9.0\tnote{g}& --\\
      \ce{CH3OH} & 5$_{1,4,0}$-4$_{1,3,0}$ & 243.9157 & 1.5 & $0\farcs22 \times 0\farcs16$ & 0.29 & $<$ 118\tnote{f} & 1.0-10.0\tnote{g} & -- \\
      \hline
      \hline
  \end{tabular}
  \begin{tablenotes}
      \item[a] Quantum numbers are formatted for \ce{H2CO} \citep[CDMS, ][]{chardon1973structure, muller2017submillimeter}, \ce{c-C3H2} \citep[CDMS, ][]{bogey1986centrifugal, vrtilek1987laboratory, lovas1992microwave} as $N_{K_a, K_c}$, for C$_2$H as $N_{J, F_1}$ \citep[CDMS, ][]{sastry1981laboratory, muller2000submillimeter, padovani2009c2h}, for SO as $N_J$ \citep[CDMS]{amano1974microwave, clark1976microwave, tiemann1982isotope, lovas1992microwave}, for \ce{CH3OH} as $N_{K_a, K_c, \nu}$ \citep[CDMS,][]{xu2008torsion, lees1968torsion, pickett1981submillimeter, sastry1984microwave, herbst1984new},  for \ce{^12CO} \citep[CDMS, ][]{winnewisser1997sub}, \ce{^13CO} \citep[CDMS, ][]{klapper2000sub, cazzoli2004precise}, \ce{C^18O} \citep[CDMS, ][]{winnewisser1985millimeter}, \ce{H^13CN} \citep[CDMS, ][]{fuchs2004high, cazzoli2005lamb, maiwald2000pure}, \ce{HC^15N} \citep[CDMS, ][]{fuchs2004high, cazzoli2005lamb}, \ce{DCN} \citep[CDMS, ][]{brunken2004sub}, \ce{H^13CO^+} \citep[CDMS, ][]{gregersen2001erratum, lattanzi2007rotational}, \ce{CS} \citep[CDMS, ][]{muller2005cologne, bogey1981millimeter, ahrens1999pure} as $J$.
      \item[b] The uncertainty does not include the 10\% absolute flux calibration.
      \item[c] Velocity range over which the spectral integration has been performed to evaluate the flux (chosen after visual inspection of the data, to include channel maps where signal is detected).
      \item[d] Limits of the integration performed after the double-Gaussian fit to the hyperfine components of the spectrum.
      \item[e] The value refers to the total flux of the blended \ce{c-C3H2} (218~GHz) lines. 
      \item[f] Upper limit on the integrated flux evaluated from the integrated intensity maps, as 3 times the standard deviation of the flux of 26 de-projected circles, taken outside the possible emitting region, with a radius of 2$\as$.
      \item[g] Velocity range over which we extracted the integrated intensity maps, from which the indicated upper limit on the line flux has been estimated. The range has been selected to match the average one from detected lines, compatible with the channel width of cubes of undetected lines.  

  \end{tablenotes}
  \end{threeparttable}
\end{table*} 

In this paper, we present ALMA band~6 line emission observations from the ALMA program \#2019.1.01619.S (PI S. Facchini): we complement the results for the short baselines, previously shown by \cite{bib:facchini2021chemical}, with the long-baseline data. We performed the following data reduction through the \texttt{CASA} software \citep{mcmullin2007casa} \texttt{v5.8}.

The observations included two spectral setups. The calibration of the lower-frequency setup is reported by \cite{law2024mapping}. We calibrated the high-frequency observations with the same procedure. These observations consist of two execution blocks (EBs) for the short spacings, and five EBs in an extended configuration. In summary, we first realigned the observations of each EB by shifting the phase center to the center of an ellipse fitted to the outer continuum ring of PDS 70. To perform the shift, we used the \texttt{CASA} tasks \texttt{fixvis} and \texttt{fixplanets}. After flagging the channels in the proximity of the lines detected by \citet{bib:facchini2021chemical} in a $\pm 15\,$km\,s$^{-1}$ range from the systemic velocity, we averaged the data into 250\,MHz channels. We then self-calibrated the continuum data, the short baselines first, and then the concatenated short and long baselines \citep[as in][]{andrews2018disk}. For the short-baseline data, we used the same solution intervals as in \citet{bib:facchini2021chemical}, whereas for data combining short and long baselines we computed and applied a phase-only solution down to 120\,s. We ran a final round of amplitude and phase self-calibration on EB-long intervals. The resulting continuum image reconstructed with the \texttt{tclean} task, using the \texttt{multiscale} deconvolver, with Briggs weighting and a \texttt{robust} parameter of 0.5 shows a root mean square (RMS) of 9.27\,$\mu$Jy. The peak signal-to-noise ratio (S/N) is 234, over a synthesized beam of $0\farcs14\times0\farcs10$ (with a PA of -85.3$^\circ$). The total flux density of the disk at the weighted frequency of 252.8\,GHz is $68.42\pm0.10\,$mJy, without accounting for 10\% absolute flux calibration uncertainties, which dominate but only result in a systematic shift. The flux uncertainty was evaluated from standard deviations of fluxes from random ellipses with a minor axis of $1\farcs5 \times \cos{(i)}$ ($i$ is the disk inclination) and PA as the disk, taken outside the emitting area (the same procedure was applied for line fluxes, see Sect.~\ref{subsec:fluxes} for more details). The gain solutions were then applied to the full spectral data. Finally, we subtracted the continuum emission using the \texttt{uvcontsub} task with a first-order polynomial. 

We imaged the different lines using the \texttt{tclean} task, based on the \texttt{CLEAN} algorithm \citep{hogbom1974aperture}, using natural weighting. We applied the \texttt{multiscale} deconvolver, with scales (0, 5, 10, 20, 30)pix, where pix indicates the pixel dimension ($0\farcs02\times0\farcs02$).
We tested the effect of different imaging procedures on the final image, such as different flux thresholds to stop the cleaning (2$\sigma$ and 3.5$\sigma$) and two different masking techniques: we built a first mask by cleaning down to 7$\sigma$, and a second one by leveraging the Keplerian rotation, using the \texttt{keplerian$\_$mask.py} tool by \citet{kepmask}, and we obtained consistent images. We fixed the geometrical parameters, $M_\ast = 0.875$ M$_\odot$, $i = 51.7^\circ$, PA$=160.4^\circ$, and $v_\mathrm{sys}=5.5$~km~s$^{-1}$ (in LSRK frame), as they were obtained by \cite{keppler2019highly}, and a distance of $113$ pc \citep{brown2018gaia}. We subsequently convolved the mask with a Gaussian kernel with a full width at half maximum (FWHM) of $0\farcs4$ to reproduce the broadening effect of the finite resolution in the data. The systematic difference in the resulting channel maps due to the different masks are within the statistical uncertainties. We performed all the following analyses on cubes imaged using the 7$\sigma$ mask.

In order to maximize the detection of the faint lines H$_2$CO (3$_{2,1}$-2$_{2,0}$), \ce{c-C3H2}, SO, and \ce{CH3OH}, we imaged their data cubes with an additional $uv$-tapering of $0\farcs2$ to increase sensitivity to extended emission.
We performed all the following analyses on the continuum subtracted cubes.

The spectral setup includes several molecular transitions, including CO isotopologs ($^{12}$CO, C$^{18}$O, $^{13}$CO), formaldehyde (H$_2$CO), cyanides (H$^{13}$CN, HC$^{15}$N, DCN), hydrocarbons (C$_2$H, \ce{c-C3H2}), S-bearing molecules (CS, SO), ions (H$^{13}$CO$^+$), and methanol (\ce{CH3OH}). In particular, we observed two molecular transitions of \ce{c-C3H2}, one at 244.2221 GHz (3$_{2,1}$-2$_{1,2}$) and another at 218.7327 GHz: the latter corresponds to the two ortho (7$_{1,6}$-7$_{0,7}$) and para (7$_{2,6}$-7$_{1,7}$) lines, which are blended. In the paper, we refer to the 3$_{2,1}$-2$_{1,2}$ transition as “\ce{c-C3H2} (244 GHz)” and to the blended one as “\ce{c-C3H2} (218 GHz).”
An overview of the analyzed transitions, corresponding beam size, RMS, and line flux is presented in Table~\ref{tab:imaging_molecules} (see also Sect.~\ref{subsec:fluxes}).

We accounted for the so-called JvM effect \citep{jorsater1995high}, following \cite{czekala2021molecules}, by rescaling the residual map by the $\epsilon$ factor, which is the ratio between areas of the CLEAN and DIRTY beams. $\epsilon$ factors are listed in Table~\ref{tab:imaging_molecules} for each line. Non-JvM corrected images show higher residuals, as they are not rescaled, resulting in an overestimated line flux, by a factor that we estimated to be $\sim$10\% for bright lines, reaching $\sim$40\% for faint lines. In this work, the analysis was performed on the JvM corrected cubes, except for undetected lines, not to overestimate the sensitivity of the observations \citep{casassus2022variable}. For the same reason, we used non-JvM corrected images in the rotational diagram analysis of the \ce{H2CO} asymmetry (see Sect.~\ref{subsec:H2CO}), as it is performed on spatial scales equal to the beam size.

\begin{figure*}[ht!]
    \includegraphics[width=1.03\textwidth, center]{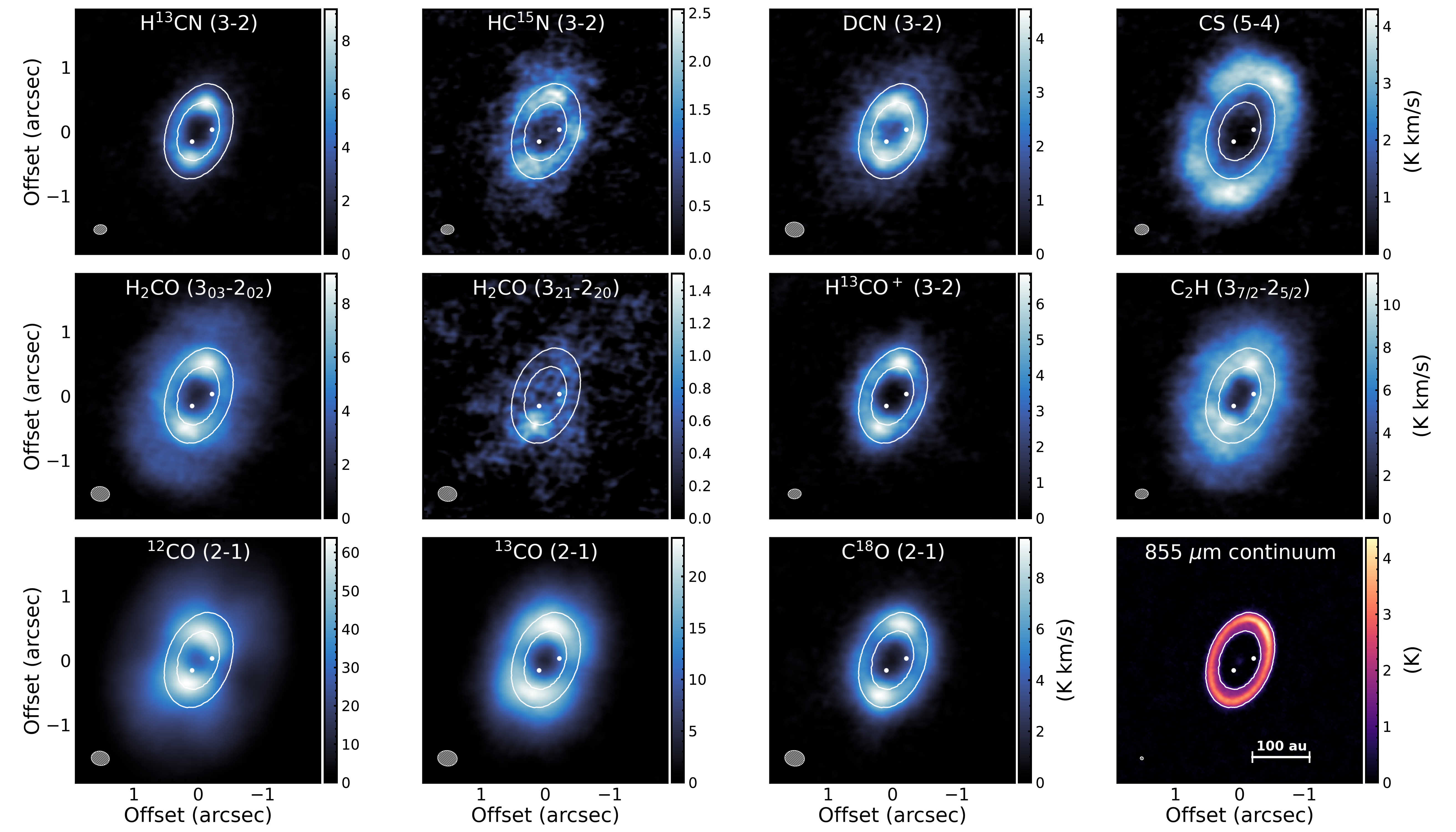}
    \caption{Integrated intensity maps of the detected lines, and of the 855$\mu$m continuum \citep{isella2019detection}, with the white contours showing the bright ring in the submillimeter continuum emission and the white dots marking the position of the two forming planets \citep{wang2021constraining}. Brightness temperatures were obtained under the assumption of Rayleigh-Jeans approximation. 
    The white ellipse at the bottom left of each panel is the beam, while the white line at the bottom right of the last panel indicates the 100 au scale.}
    \label{fig:M0}
\end{figure*} 

\section{Boosting faint lines detection} \label{sec:boost}
In this work, we applied boosting techniques both in the image and in the visibility plane, to the weak lines included in the dataset: H$_2$CO (3$_{2,1}$-2$_{2,0}$), \ce{c-C3H2} (244 GHz), \ce{c-C3H2} (218 GHz), SO (6$_7$-5$_6$), and SO (6$_6$-5$_5$).
Techniques in the image plane leverage the disk rotation pattern, to stack the spatially integrated line spectrum, by shifting back the spectrum along each spatial pixel to the systemic velocity, and boost the S/N \citep[][and Fig.~\ref{fig:spectra}]{yen2016stacking}. This method has been implemented in the \texttt{GoFish} package \citep{teague2019gofish}, and allows one to reconstruct a shifted spectrum by correcting for the Keplerian rotation (see also Appendix~\ref{appendix:int_spectra}).

On the other hand, these techniques are sensitive to imaging artifacts, spatial covariance on the beam scale and/or spectral information loss due to spectral masks \citep{loomis2018visible}. Directly applying boosting techniques in the visibility plane could prevent limitations linked to the image reconstruction. In this context, multiple lines of the same molecule can be stacked in the $uv$-plane \citep{walsh2016first}, or a matched filter analysis can be applied by spanning the sampled visibilities of the faint line with a model filter \citep{loomis2018visible}.

\subsection{Detection of \texorpdfstring{$H_2CO \, (3_{2,1}-2_{2,0}$)}{H2CO (321-220)}} \label{subsec:H2CO}

Two molecular lines of H$_2$CO are included in the observed spectral window: the bright ring-shaped 3$_{0,3}$-2$_{0,2}$ line, which is detected at a high S/N (see Fig.~\ref{fig:M0}) and the faint and asymmetric 3$_{2,1}$-2$_{2,0}$ transition. For the case of the faint H$_2$CO 3$_{2,1}$-2$_{2,0}$ line, the Keplerian mask and natural weighting during the CLEANing procedure were enough to retrieve the Keplerian pattern in the channel maps. The emission distribution is represented in the integrated intensity map in Fig.~\ref{fig:M0}, where the signal is spatially resolved and asymmetric. 

To confirm the robustness of the detection, we used the \texttt{GoFish} package \citep{teague2019gofish} to extract the Keplerian shifted spatially integrated spectrum within a circle of 2$\as$ of radius, which is represented in the top right plot in Fig.~\ref{fig:spectra}. We also obtained the Keplerian shifted spectrum as a function of radius, in order to highlight the spatial location of the emitted signal, as it is represented in the bottom right panel in Fig.~\ref{fig:spectra}. In this case, image-plane techniques were enough to confirm the detection, with an S/N of 9.5 (evaluated from the integrated flux, see Sect. \ref{subsec:fluxes}), which confirms the marginal detection of $4.4\sigma$ presented by \cite{bib:facchini2021chemical}.

\subsection{Detection of c-\texorpdfstring{$\,C_3H_2$}{C3H2}} \label{subsec:C3H2}

\begin{figure*}[ht!]
    \centering
    \includegraphics[width=1.15\textwidth, center]{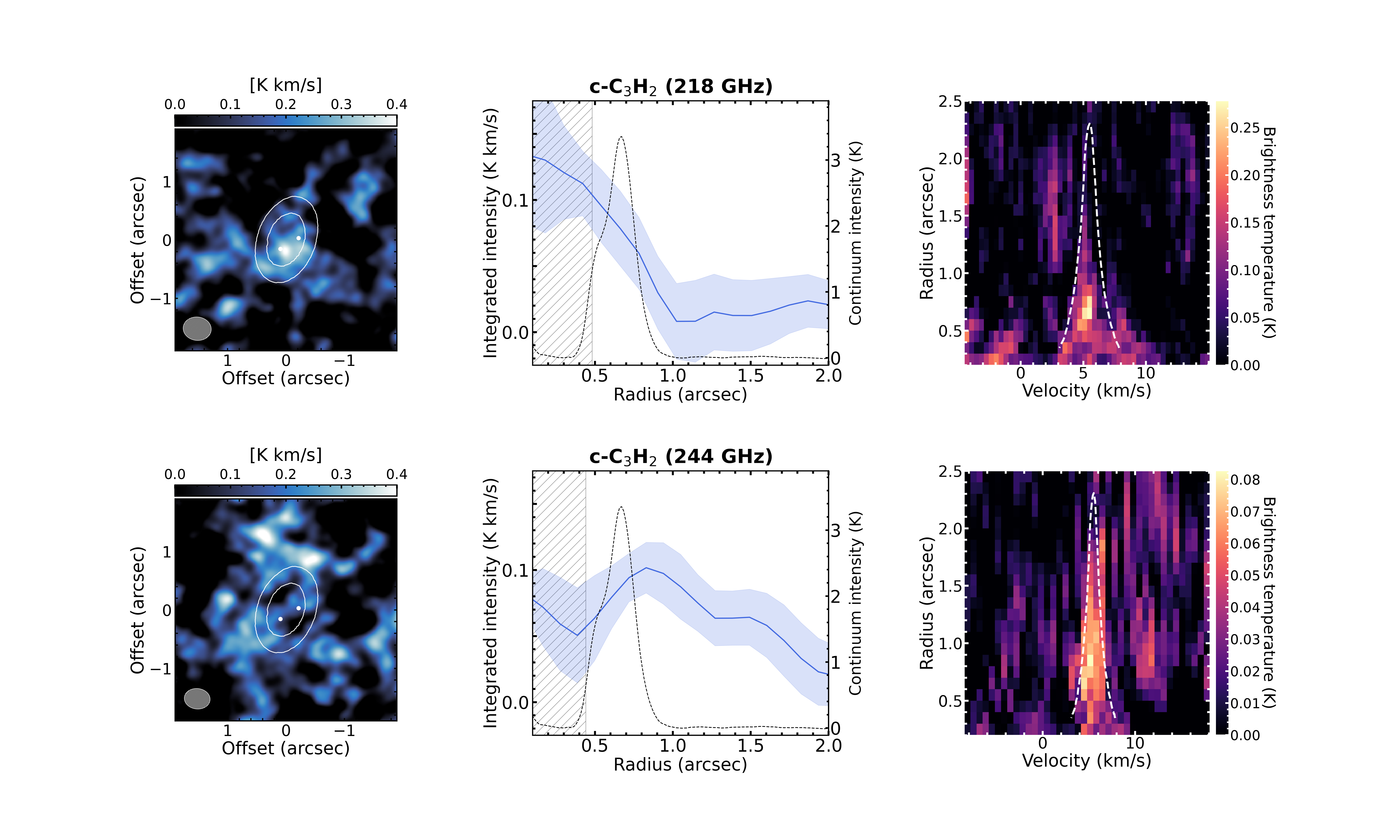}
    \caption{Detection of the \ce{c-C3H2} transitions at 218 GHz (top row) and at 244 GHz (bottom row). Panels in the left column show the integrated intensity maps obtained from the imaged data cubes with natural weighting and an additional $uv$-tapering. The white contours show the band 7 submillimeter continuum bright ring, and the two white dots indicate the location of the two gas giants \citep{wang2021constraining}, while the white ellipses on the bottom left are the beams. Panels in the central column represent the radial profiles of the azimuthally averaged integrated intensity, obtained as described in Sect.~\ref{subsec:m0}, for \ce{c-C3H2} transitions (blue line), and for the 855$\mu$m continuum emission \citep[dashed black line,][]{isella2019detection}. The hatched gray region indicates the beam major axis. Panels in the right column show the Keplerian shifted line spectra, in brightness temperature, as a function of radius, for each $0\farcs1$ wide annulus in which the disk has been divided. The dashed white lines are the 3K level of the $^{12}$CO.}
    \label{fig:C3H2}
\end{figure*}

\begin{figure*}[ht!]
    \centering \includegraphics[width=1.0\textwidth, center]{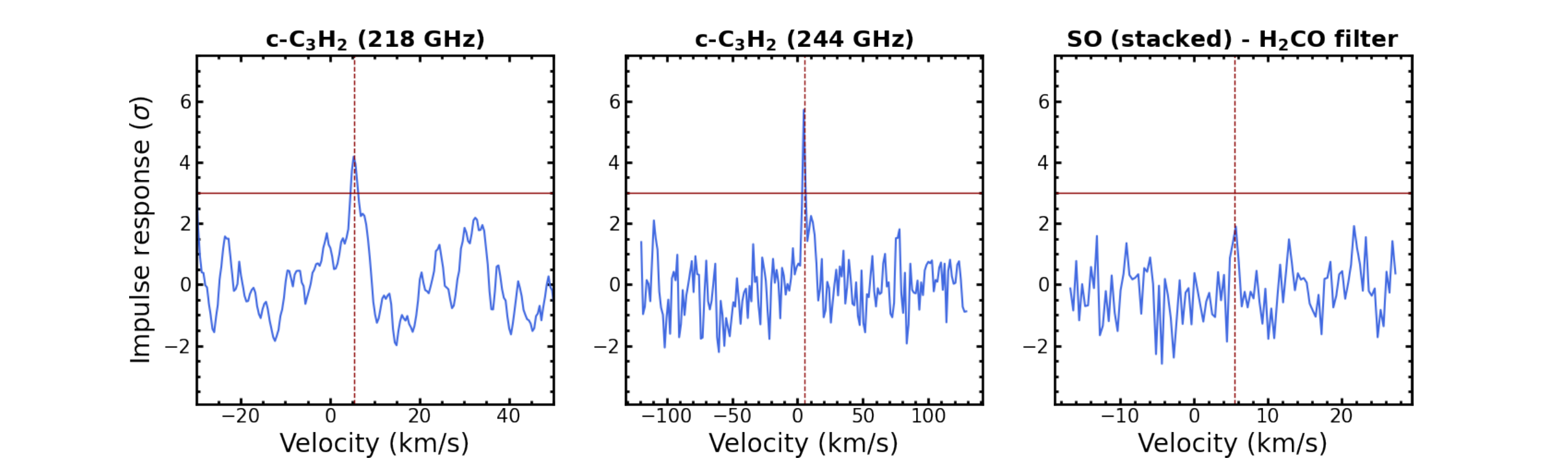}
    \caption{Filter response as a function of velocity, as we obtained from the matched filter analysis performed on the \ce{c-C3H2} (218 GHz, left panel), and the C$_3$H$_2$ (244 GHz, middle panel) transitions, with a Keplerian mask as filter: the vertical dashed red line indicates the systemic velocity and the position of the peaks corresponding to the two \ce{c-C3H2} lines. Right panel: Impulse response from the matched filter analysis performed on the SO stacked visibilities, using the clean image of the H$_2$CO (3$_{0,3}$-2$_{0,2}$) line as a filter. The horizontal red line shows the 3$\sigma$ level.}
    \label{fig:visible_faint}
\end{figure*}

We applied boosting techniques in the image plane to the two lines of the \ce{c-C3H2} molecule: we imaged the two lines with natural weighting and an additional $uv$-tapering of $0\farcs2$, along with a Keplerian mask, as we described in Sect.~\ref{sec:data}.  The emission is at least marginally detected in the channel maps of both lines, as it can be seen from the top and middle panels in Fig.~\ref{fig:channel_maps}, where we highlight the 3$\sigma$ contours in white. From the integrated intensity maps (see first column in Fig.~\ref{fig:C3H2}), we also extracted the radial profile of the integrated intensity. The results are presented in the middle column of Fig.~\ref{fig:C3H2}.

To assess the significance of the signal, we extracted the Keplerian shifted spectra as a function of radius \citep{teague2019gofish}, which are shown in the right column of Fig.~\ref{fig:C3H2}, with the dashed white lines defining the 3K level from the $^{12}$CO line. The two shifted spectra are centered on the systemic velocity, supporting the detection. As is also evident from the integrated intensity maps (left column Fig.~\ref{fig:C3H2}), the two lines show a different emission morphology, with the 218 GHz ones being more compact. This could be due to the fact that the two transitions have different upper energies, and are sensitive to different environmental conditions, resulting in different emitting regions. This can be critical when extracting the rotational diagram analysis, as we discuss further in Sect.~\ref{subsec:faint_lines_C3H2}. 

In summary, \ce{c-C3H2} 218 GHz lines are marginally detected with an S/N of 3.3 (undetected in the short-baseline data, \citealt{bib:facchini2021chemical}), while the \ce{c-C3H2} 244 GHz transition is detected with an S/N of 8 (marginally detected by \citealt{bib:facchini2021chemical} with an S/N of 4.3). Signal-to-noise ratios were derived from disk-integrated fluxes, extracted after applying boosting techniques (see Sect.~\ref{subsec:fluxes}).

To  confirm the detection, we applied the matched filter analysis directly in the $uv$-plane, using the \texttt{visible} package \citep{loomis2018visible}. We built a model filter from a Keplerian mask using the \texttt{Keplerian$\_$mask.py} tool developed by \cite{kepmask}, assuming the geometrical and line parameters that we applied in the imaging, but varying the radial extent, in order to find the mask that maximizes the filter response, when spanning the kernel over the sampled visibilities. The best results for the filter responses to the \ce{c-C3H2} lines use a mask with an inner radius of $0\as$ and an outer radius of $1\as$ (218 GHz) and $2\as$ (244 GHz). The corresponding filter responses are shown in the left and middle panels of Fig.~\ref{fig:visible_faint}, as a function of velocity. We applied the analysis after running the ms tables through the \texttt{cvel2} task, which corrects for Doppler shifts throughout the time of the observation, and transforms the reference frame from TOPO to LSRK, knowing the rest frequencies of the targeted lines. We therefore expect the peak of the filter response to be at the systemic velocity, as it is marked by the vertical red lines for the targeted faint \ce{c-C3H2} lines, in Fig.~\ref{fig:visible_faint}. The matched filter response peaks at the systemic velocity (5.5~km~s$^{-1}$), and thus confirms the detection of the two lines, with a peak filter response of 5.7$\sigma$ (for the 244 GHz line) and of 4.2$\sigma$ (for the 218 GHz lines).

\subsection{\texorpdfstring{$uv$}{uv}-stacking of SO}
\label{subsec:stacking_SO}

The imaging techniques we applied for the other weak molecular transitions — that is, using natural weighting, and an additional $uv$-tapering, or Keplerian shifting — did not lead to a detection when applied to the two SO 6$_7$-5$_6$ and SO 6$_6$-5$_5$ lines. We therefore applied a stacking technique in the $uv$-plane, by combining the two individually undetected lines, as it has been previously presented by \cite{walsh2016first} for the case of the methanol detection in the TW Hya disk. As the two lines have different rest frequencies, we re-gridded the data using \texttt{cvel2}, such that the central frequency of each line refers to the systemic velocity, and they can subsequently be correctly stacked. We then combined the two measurement sets using the \texttt{CASA} task \texttt{concat}: Fig.~\ref{fig:test_uv_stacking} shows a test of the $uv$-stacking performed on two molecular lines of H$_2$CO that are also individually detected, where the integrated intensity profile of the stacked cube (pink) lies between the higher and the lower profiles, representing the integrated intensity profiles of the bright (orange) and the faint (blue) lines, respectively. 

We then imaged the stacked SO cube, using the same parameters applied for the re-gridding in \texttt{cvel2} (channel width, spectral extent, starting channel). As we expected from the stacking procedure, the stacked data cube shows an RMS $\sim$~$\sqrt{2}$ times lower than the original ones. We did not detect any signal in the SO channel maps of the $uv$-stacked cube above the 3$\sigma$ level. 
The right panel in Fig.~\ref{fig:SO_uv_stacking} shows the Keplerian shifted spectra as a function of radius, for the original undetected lines (left and middle panels), and for the $uv$-stacked cube (right panel), without any clear boosted emission linked to the expected Keplerian pattern of the disk. 
The bottom panel in Fig.~\ref{fig:channel_maps} shows the channel maps obtained from the cleaned $uv$-stacked data cube, where the white contours show the 3$\sigma$ level. The only marginal detection originates from the outer disk at 4.10~km~s$^{-1}$ (see Sect.~\ref{subsec:faint_lines_SO} for a detailed discussion).

We applied the matched filter analysis to the stacked visibilities, testing various model filters, built from Keplerian masks, or from the emission of other molecules. The strongest filter response is obtained using \ce{H2CO} 3$_{0,3}$-2$_{0,2}$ as a filter, as it shows extended emission, and is thus more sensitive to the 3$\sigma$ feature observed in the outer disk. However, the result from the matched filter analysis is not significant, since it is below the 3$\sigma$ threshold, as is shown in Fig.~\ref{fig:visible_faint}.

\begin{figure*}[ht!]
    \includegraphics[width=1.1\textwidth, center]{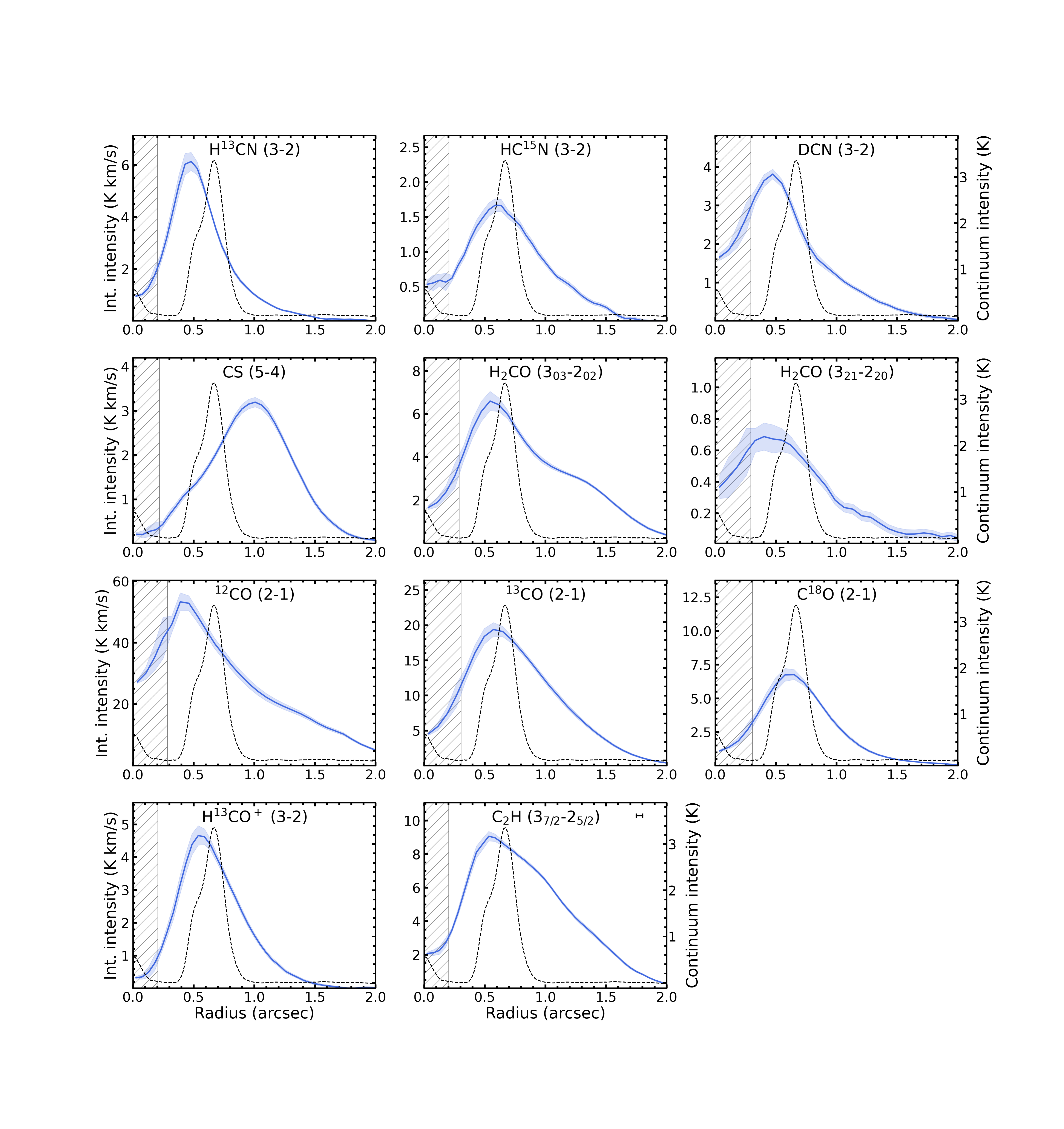}
    \vspace{-2cm}
    \caption{Azimuthally averaged radial profiles of the integrated intensity for the analyzed molecular transitions.
    The ribbons show the standard deviation across each annulus, divided by the square root of the number of independent beams. The FWHM of the synthesized beam is highlighted by the hatched region, while the beam size of the continuum image is indicated on the top-right corner of the last panel. The dashed black line shows the azimuthally averaged radial profile of the 855$\mu$m continuum emission \citep{isella2019detection}.}
    \label{fig:profiles}
\end{figure*} 

\section{Overview of the observed molecules} 
\label{sec:results_molecules}

In this section, we present an overview of the analyzes we applied to reconstruct the emission morphology and line fluxes from the high resolution observations (see \citealt{bib:facchini2021chemical} for reference on the methodologies).

\subsection{Disk-integrated spectra} \label{subsec:spectra}
After reconstructing the emission distribution of the various molecular transitions, we extracted line spectra, by applying the stacking method in the image plane, presented in Sect.~\ref{sec:boost}.
We used the \texttt{GoFish} package \citep{teague2019gofish}, following \cite{teague2016measuring}, under the assumption of an axisymmetric Keplerian disk and from the geometrical parameters and the star mass presented in Sect.~\ref{sec:data}.  We assumed midplane emission for all molecules except for the most elevated $^{12}$CO and $^{13}$CO emission. For the latters, we included a tapered power law prescription of the emitting layer height $z(r)$:
\begin{equation}\label{eq:z}
    z(r) =
    \begin{cases}
     z_0 \times \left( \dfrac{r -r_\mathrm{cavity}}{1\as}\right)^\phi \times \exp{\left( - \left[ \dfrac{r}{r_\mathrm{taper}}\right]^\psi {}\right)} & r > r_\mathrm{cavity} \\
     0 & r \leq r_\mathrm{cavity}, 
    \end{cases}
\end{equation}
with $z_0=0\farcs41$, $r_\mathrm{cavity}=0\farcs37$, $\phi=0.51$, $r_\mathrm{taper}=1\farcs27$, $\psi=5.74$ for \ce{^12CO}, and $z_0=0\farcs41$, $r_\mathrm{cavity}=0\farcs28$, $\phi=1.29$, $r_\mathrm{taper}=0\farcs74$, $\psi=1.61$ for \ce{^13CO} \citep{law2024mapping}. We then extracted the integrated spectrum by spatially integrating the shifted spectra over a de-projected circle of radius 3$\as$ for the lines with the most extended emission ($^{12}$CO and $^{13}$CO), and 2$\as$ for the other ones. Radial extents are chosen to be large enough not to exclude extended emission, but not to include excessive noise in the calculation. An example of the extracted disk-integrated spectra is shown in the top row of Fig.~\ref{fig:spectra}.

\subsection{Integrated intensity maps and radial profiles} \label{subsec:m0} 

We extracted the integrated intensity maps of the detected lines by spectrally integrating the JvM-corrected data cubes with the \texttt{bettermoments} tool \citep{teague2019bettermoments}. We collapsed the cubes over the velocity ranges listed in Table~\ref{tab:imaging_molecules}, chosen after visual inspection of the data to include all channel maps showing signal, without applying any additional mask. Figure~\ref{fig:M0} shows the integrated intensity maps of the detected lines.
The position of the two planets is marked with white dots \citep{wang2021constraining}, while the white contours show the bright ring in the ALMA band~7 submillimeter continuum emission \citep{isella2019detection}. The azimuthal dark feature in the CS map is a channelization effect due to low spectral resolution. All molecules have a structured emission morphology, with a ringed-shape emission, not always colocated with the bright ring in the submillimeter continuum.
           
We extracted the radial profiles from the integrated intensity maps shown in Fig.~\ref{fig:M0}, using the \texttt{GoFish} package \citep{teague2019gofish}, by dividing each map into annuli with a radial extent of 1/4 of the beam major axis, and taking the azimuthal average of the integrated intensity for each annulus (see \citealt{bib:facchini2021chemical} for more details). The result is shown for each line in Fig.~\ref{fig:profiles}: the integrated intensity is expressed in integrated brightness temperature, assuming the Rayleigh-Jeans approximation. Uncertainties are obtained by taking the standard deviation along each annulus, divided by the square root of the number of independent beams in the annulus. For radii smaller than the beam major axis we took only the standard deviation along the annuli. The uncertainties may be underestimated, especially in the inner regions, inside the beam FWHM, as it does not account for the 2D covariance introduced by the beam. 

The azimuthal average is performed after de-projecting the emission, by using the geometrical parameters presented in Sect.~\ref{sec:data}, assuming the surfaces from Eq.~\ref{eq:z} for \ce{^12CO} and \ce{^13CO}, and midplane origin for the other molecules. All molecules in Fig.~\ref{fig:profiles} have a peaked integrated intensity profile, with the \ce{^12CO}, \ce{H^13CN}, DCN, and \ce{H2CO} (3$_{2,1}$-2$_{2,0}$) lines showing the innermost peak, inside the millimeter dust peak. Each molecule presents an inner cavity, with \ce{H^13CN}, \ce{H^13CO^+}, and \ce{C2H} showing the steepest radial integrated intensity gradient, while \ce{H2CO}, \ce{C2H}, and \ce{^12CO} also show an outer shoulder of emission. Most of the molecules show a peak of the emission inside the submillimeter continuum peak, which we discuss further in Sect.~\ref{subsec:discuss_lines}, in light of the efficiently irradiated cavity wall.

\subsection{Line fluxes} \label{subsec:fluxes}

The line flux of each transition is obtained from the integrated intensity maps in Fig.~\ref{fig:M0}, by spatially integrating the maps over de-projected circles of the same extent used for the integrated spectra, in Sect.~\ref{subsec:spectra}. This procedure has been applied to all transitions except for the faintest two lines of \ce{c-C3H2}. In the latter case, we extracted the line flux from the shifted spectra as a function of radius (see right panel in Fig.~\ref{fig:C3H2}). We then applied a mask based on the \ce{^12CO} 3\,K brightness temperature level, which is marked by the dashed white line in the right panel in Fig.~\ref{fig:C3H2}. We chose to define such a mask to select a region large enough to include all the line flux, but with a different radial extent for each velocity channel, to avoid including excessive noise. The 3\,K level from \ce{^12CO} is a good indicator of the extent of the emission, expected from a Keplerian disk (see also bottom-left panel in Fig.~\ref{fig:spectra}). This procedure is similar to applying a Keplerian mask when extracting the integrated intensity map, but it prevents one from producing artifacts when creating the image. We tested this procedure on the two weakest lines, in order to robustly extract their line flux: their shifted spectra as a function of radius and the mask are shown in the right panel in Fig.~\ref{fig:C3H2}.

Table~\ref{tab:imaging_molecules} lists line fluxes and associated uncertainties for each line. The uncertainty is evaluated as the standard deviation of the flux measured in 26 de-projected circles, outside the emitting area of the integrated intensity maps, with the same radius used to extract the flux. We took the maximum number of circles that was possible to place on the image without overlap. This procedure has been performed on images which are not primary beam corrected, to ensure uniform noise. For the case of \ce{c-C3H2}, we extracted the uncertainty associated with the line flux from the shifted spectra as a function of radius, by taking the standard deviation of 26 flux estimates, which we evaluated inside the 3K mask, randomly applied outside the spectral range used to evaluate the line flux.

For the undetected SO (6$_{7}$-5$_{6}$), SO (6$_{6}$-5$_{5}$), and \ce{CH3OH} (5$_{1,4,0}$-4$_{1,3,0}$) lines, we provide an upper limit on the flux, evaluated from the integrated intensity maps, as 3 times the standard deviation of the
flux of 26 de-projected circles, taken outside the possible emitting region. For the case of the C$_2$H (3$_{7/2}$-2$_{5/2}$) transition, as the two hyperfine components are spectrally resolved, we extracted the integrated flux of the individual components by fitting a double-Gaussian to the shifted-stacked spectrum (blue line in the top-middle panel in Fig.~\ref{fig:spectra}), under the assumption of two optically thin components. The uncertainty on the flux estimate has been extracted as for the other lines from the integrated intensity map.

\subsection{Disk-averaged column densities} \label{subsec:N_av}
\begin{table*}
\begin{center}
\renewcommand{\arraystretch}{1.3}
\caption{Line parameters for the various species for which we estimated the column density.} 
\label{tab:line_par}
\begin{threeparttable}
\begin{tabular}{ c c c c c c c c c}
    \hline
    \hline
    Species & Transition\tnote{a} & $E_\mathrm{u}$ & log$_{10}$(A$_\mathrm{ul}$s$^{-1}$) & $g_\mathrm{u}$ & Q(18.75K) & Q(37.50K) & Q(75K) & $N^{\mathrm{thin}}_\mathrm{av}$ \tnote{b} \\
    &  & [K] & & & & & & [cm$^{-2}$] \\
    \hline
    C$^{18}$O & 2-1 & 15.81 & -6.22 & 5 & 7.461 & 14.57 & 28.81 & 1.1$_{-0.1}^{+0.4}$ $\times$10$^{15}$ \\
    H$_2$CO & 3$_{0,3}$-2$_{0,2}$ & 21.0 & -3.55 & 7 & 44.68 & 128.7 & 361.7 & 2.3$_{-0.6}^{+1.5}$ $\times$10$^{13}$\\
    H$_2$CO & 3$_{2,1}$-2$_{2,0}$ & 68.1 & -3.80 & 7 & 44.68 & 128.7 & 361.7 & 1.6$_{-0.4}^{+1.3}$ $\times$10$^{13}$\\
    C$_2$H & $3_{7/2, 4}$-$2_{5/2, 3}$ & 25.16 & -4.3 & 9 & 37.14 & 72.91 & 144.7 & 7.5$_{-0.1}^{+1.5}$ $\times$10$^{13}$ \\
    C$_2$H & $3_{7/2, 3}$-$2_{5/2, 2}$ & 25.15 & -4.3 & 7 & 37.14 & 72.91 & 144.7 & 7.7$_{-0.1}^{+1.6}$ $\times$10$^{13}$\\
    H$^{13}$CN & 3-2 & 24.86 & -3.11 & 21 & 28.17 & 55.31 & 109.6 & 7.8$_{-0.2}^{+1.7}$ $\times$10$^{11}$\\
    HC$^{15}$N & 3-2  & 24.78 & -3.11 & 7 & 9.421 & 18.50 & 36.66 & 3.9$_{-0.2}^{+0.9}$ $\times$10$^{11}$\\
    DCN & 3-2 & 20.85 & -3.33 & 21 & 33.39 & 65.75 & 130.5 &  8.7$_{-0.6}^{+2.5}$ $\times$10$^{11}$ \\
    H$^{13}$CO$^+$ & 3-2 & 24.98 & -2.87 & 7 & 9.348 & 18.35 & 36.37 & 4.7$_{-0.1}^{+0.9}$ $\times$10$^{11}$\\
    CS & 5-4 & 35.27 & -3.53 & 11 & 16.29 & 32.24 & 64.15 & 3.9$_{-0.1}^{+0.9}$ $\times$10$^{12}$\\
    \ce{c-C3H2} & ortho 3$_{21}$-2$_{12}$ & 18.17 & -4.23 & 21 & 201.8 & 566.9 & 1598 & 5.9$_{-2.1}^{+5.0}$ $\times$10$^{12}$\\
    \ce{c-C3H2} & ortho 7$_{1,6}$-7$_{0,7}$ & 61.17 & -4.05 & 45 & 201.8\tnote{c} & 566.9\tnote{c} & 1598\tnote{c} & 2.0$_{-0.7}^{+1.9}$ $\times$10$^{12}$\\
    \ce{c-C3H2} & para 7$_{2,6}$-7$_{1,7}$ & 61.17 & -4.05 & 15 & 201.8\tnote{c} & 566.9\tnote{c} & 1598\tnote{c} & 3.6$_{-1.2}^{+3.3}$ $\times$10$^{12}$\\
    SO & 6$_7$-5$_6$ & 47.55 & -3.64 & 15 & 38.88 & 90.34 & 197.51 & $<$3.0$\times$10$^{12}$ \tnote{d}\\
    SO & 6$_6$-5$_5$ & 56.5 & -3.67 & 13 & 38.88 & 90.34 & 197.51 & $<$8.0$\times$10$^{12}$ \tnote{d}\\
    \ce{CH3OH} & 5$_{1,4,0}$-4$_{1,3,0}$ & 49.66 & -4.22 & 44 & 275.0 & 920.9 & 2924 & $<$4$\times$10$^{13}$ \tnote{d}\\
    \hline
    \hline
\end{tabular}
\begin{tablenotes}
     \item[a] See Table~\ref{tab:imaging_molecules} for quantum numbers formatting.
     \item[b] Average column density within the integration area of the flux estimate, for an excitation temperature of 30K. The uncertainties on the column density include the propagated uncertainty on the flux and the systematic uncertainty on the excitation temperature.
     \item[c] The calculation of the partition functions includes all lines of both ortho and para isomers.
     \item[d] For the undetected lines we present an upper limit on the column density from the upper limit on the flux.
\end{tablenotes}
\end{threeparttable}
\end{center}
\end{table*} 

\begin{figure*}[ht!]
\centering
    \includegraphics[width=0.98\textwidth]{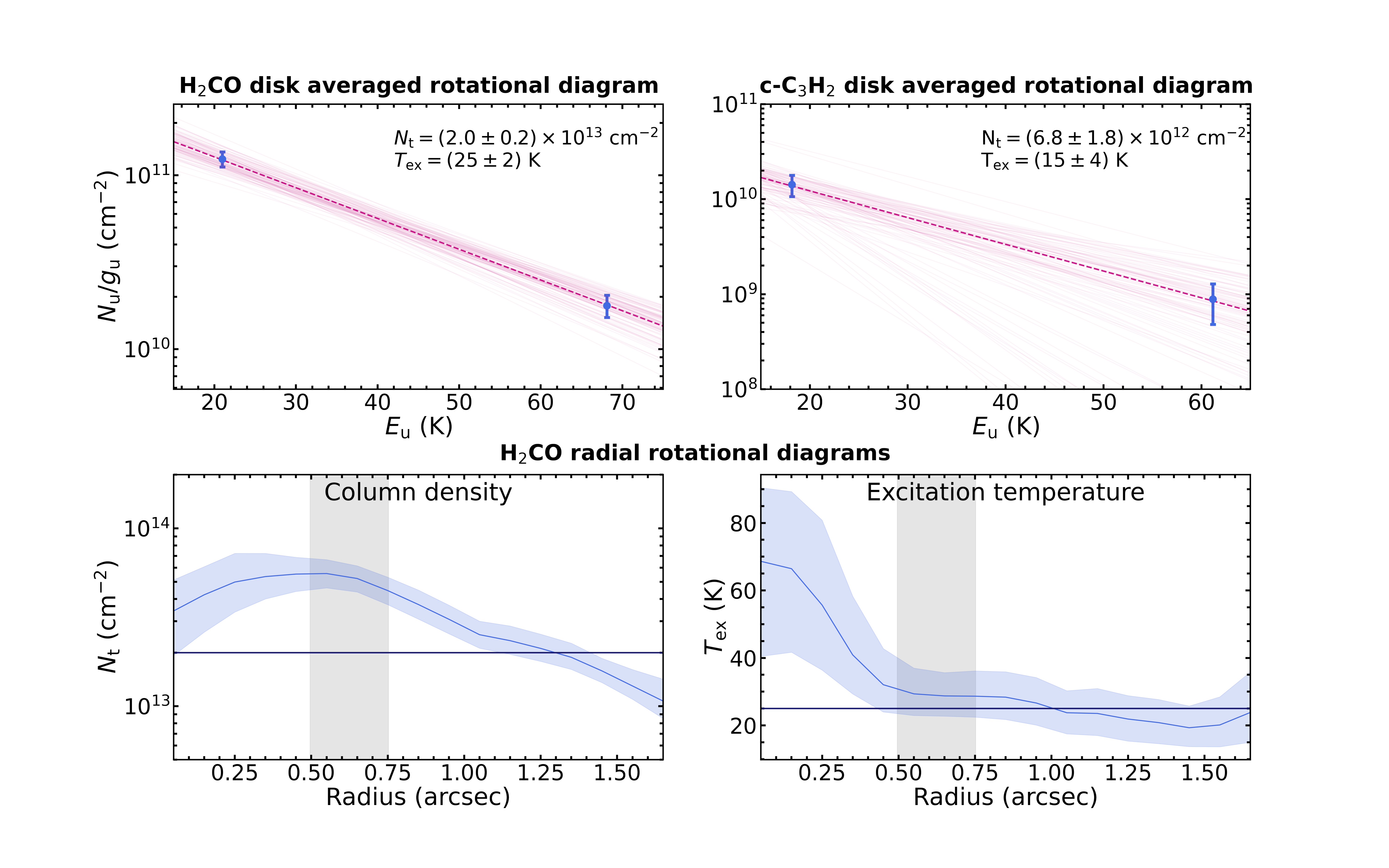}
    \caption{Rotational diagram analysis of \ce{H2CO} and \ce{c-C3H2}. First row: Disk-averaged rotational diagram analysis applied to the two lines of the H$_2$CO and \ce{c-C3H2} molecules, respectively. The average excitation temperatures and column densities are indicated on the top of the two panels.  The dashed lines correspond to the best fit results for $N_\mathrm{t}$ and $T_\mathrm{ex}$, while the solid lines are 100 random results from the posterior distribution of the \texttt{emcee} fit. Bottom row: Radial profiles of the total column density (left panel) and of the excitation temperature (right panel) from H$_2$CO rotational diagram analysis, applied to $0\farcs1$ wide annuli. The solid lines show the median of the posterior distribution (which is representative of the best fit as the posterior distribution is well described by a Gaussian) from the \texttt{emcee} fit, for $N_\mathrm{t}$ and $T_\mathrm{ex}$, while the ribbons represent the 16th and 84th percentiles. The horizontal dark blue line shows the results from the disk-average rotational diagram (from top left panel). Gray regions mark the radial range of the expected location of the CO snowline \citep{law2024mapping}.}
    \label{fig:rot_diag}
\end{figure*}

From the flux estimates listed in Table~\ref{tab:imaging_molecules}, we extracted the disk averaged molecular column density for all molecules (listed in Table~\ref{tab:line_par}), except for \ce{^12CO} and \ce{^13CO}, which are expected to be optically thick. We assumed local thermodynamic equilibrium (LTE), and optically thin emission, following \cite{bib:facchini2021chemical}. LTE is a reasonable assumption in disks, as density becomes high enough at the midplane: for the case of PDS~70 the midplane is reaching a density of $\sim$\,$10^8$\,cm$^{-3}$ \citep{portilla2023constraining}. For the molecular transitions and the excitation temperatures (see below) we are considering, critical densities are typically between $5\times 10^5$\,cm$^{-3}$ and $5\times10^6$\,cm$^{-3}$ \citep{shirley2015}: the LTE approximation is therefore generally valid for an emitting layer $z/r \lesssim 0.25$ \citep{portilla2023constraining}. The optically thin assumption is expected to be a good approximation for most of the considered molecules: \cite{law2024mapping} showed (see Fig.~12) that \ce{C^18O} is becoming marginally optically thick at its peak at the location of the bright millimeter dust ring. If we consider the peak of \ce{C^18O} in Fig.~\ref{fig:profiles} as benchmark, we see that all the molecules are below the threshold, except for \ce{^12CO} and \ce{^13CO}, for which we did not derive the column density. We took \ce{C^18O} as reference, as it is expected to emit from the lowest layer of all molecular transitions discussed in this work, with the other molecules having a kinetic temperature $T_\mathrm{kin} \gtrsim T_\mathrm{kin}$(\ce{C^18O}), and therefore a lower optical depth than the \ce{C^18O} (2-1) transition, if their brightness temperature is lower than the \ce{C^18O} one. Both \ce{C2H} and \ce{H2CO} are becoming marginally optically thick, but only at their peak: we therefore included the two molecules to extract the column density, which needs to be taken as lower limit where the molecule is becoming marginally optically thick. We used the line parameters $E_\mathrm{u}$ (upper state energy), $g_\mathrm{u}$ (upper state degeneracy), $Q(T)$ (partition function), and $A_\mathrm{ul}$ (Einstein coefficient) listed in Table~\ref{tab:line_par}, which we retrieved from the CDMS database \citep{pickett1998submillimeter, muller2001cologne}. We evaluated the partition functions by interpolating with a cubic spline, over tabulated values. 

We extracted the average molecular column density for each line, inside the de-projected circle where we evaluated the line fluxes: the results are listed in Table~\ref{tab:line_par}, for an excitation temperature of 30\,K. The errors include the propagation of the statistical uncertainty over the line flux and the systematic uncertainty from the unknown excitation temperature, which we varied between 20\,K and 50\,K. This range is consistent with the excitation temperature range extracted for \ce{H2CO} (see following discussion). We therefore evaluated the column density ${N_\mathrm{i}}$ at different excitation temperatures within the assumed range, and the related statistical uncertainty propagated from the flux uncertainty ${\mathrm{d}N_\mathrm{i}}$. To evaluate the final uncertainty over the disk averaged column density, we took the largest value of $N_\mathrm{i}+\mathrm{d}N_\mathrm{i}$ and the lowest value of $N_\mathrm{i}-\mathrm{d}N_\mathrm{i}$. We did not include the 10\% absolute flux calibration uncertainty, as it would result only in a systematic shift. For the undetected lines, we provide an upper limit on the average molecular column density from the upper limit on the line flux (see Table~\ref{tab:imaging_molecules}).

If more than one line is detected for a certain molecule, both the excitation temperature $T_\mathrm{ex}$ and the total molecular column density $N_\mathrm{t}$ can be extracted by leveraging a rotational diagram analysis \citep{goldsmith1999population}. We applied this methodology to the two lines of the H$_2$CO and the \ce{c-C3H2} molecules respectively. For the latter, we took into account the ortho- and para- spin isomers of \ce{c-C3H2}, presenting a line at the same rest frequency (218.73 GHz) and with the same line parameters, except for the upper state degeneracy ($g_\mathrm{u}=45$ for ortho and $g_\mathrm{u}=15$ for para, assuming an ortho-to-para ratio of 3:1, see Table~\ref{tab:line_par}). We extracted the disk averaged rotational diagram inside a de-projected circular area, with a radius of 2$\as$ for the more extended \ce{H2CO} and of $1\farcs5$ for the more compact \ce{c-C3H2}, obtaining $T_\mathrm{ex}=(25 \pm 2)$~K and $N_\mathrm{t}=(2.0\pm 0.2)\times10^{13}$~cm$^{-2}$ for \ce{H2CO}, $T_\mathrm{ex}= (15 \pm 4)$~K and $N_\mathrm{t}= (6.8 \pm 1.8)\times10^{12}$~cm$^{-2}$ for \ce{c-C3H2}, as it is shown in the two top panels of Fig.~\ref{fig:rot_diag}. The radius of $1\farcs5$ for the case of \ce{c-C3H2} has been chosen to balance the more extended line at 244 GHz and the more compact one at 218 GHz, as it is visible from Fig.~\ref{fig:C3H2}. For \ce{c-C3H2} we also included the 10\% absolute flux calibration uncertainty (sum in quadrature with flux statistical uncertainty), as the two lines are in different spectral setups, while we did not include it for \ce{H2CO}, as the two lines are in the same spectral setup, with absolute flux calibration uncertainty possibly resulting in the same systematic shift. We explored the posterior distribution through the MCMC method implemented in the \texttt{emcee} package \citep{foreman2013emcee}, with 128 walkers, 5000 burn-in steps, and 500 steps. Error bars refer to the 16th and 84th percentiles of the posterior distribution. We considered a uniform prior with $T_\mathrm{ex}\in(0,50)$\,K, and $N_\mathrm{t}\in(10^{11}, 10^{14})$\,cm$^{-2}$. Our result for the \ce{H2CO} molecule is consistent with what \cite{bib:facchini2021chemical} presented, while they did not perform the rotational diagram analysis on \ce{c-C3H2}, as the faint lines at 218~GHz were undetected.

\subsection{Radial profiles of column density} \label{subsec:N_rad}

\begin{figure*}[ht!]
    \includegraphics[scale=0.13, center]{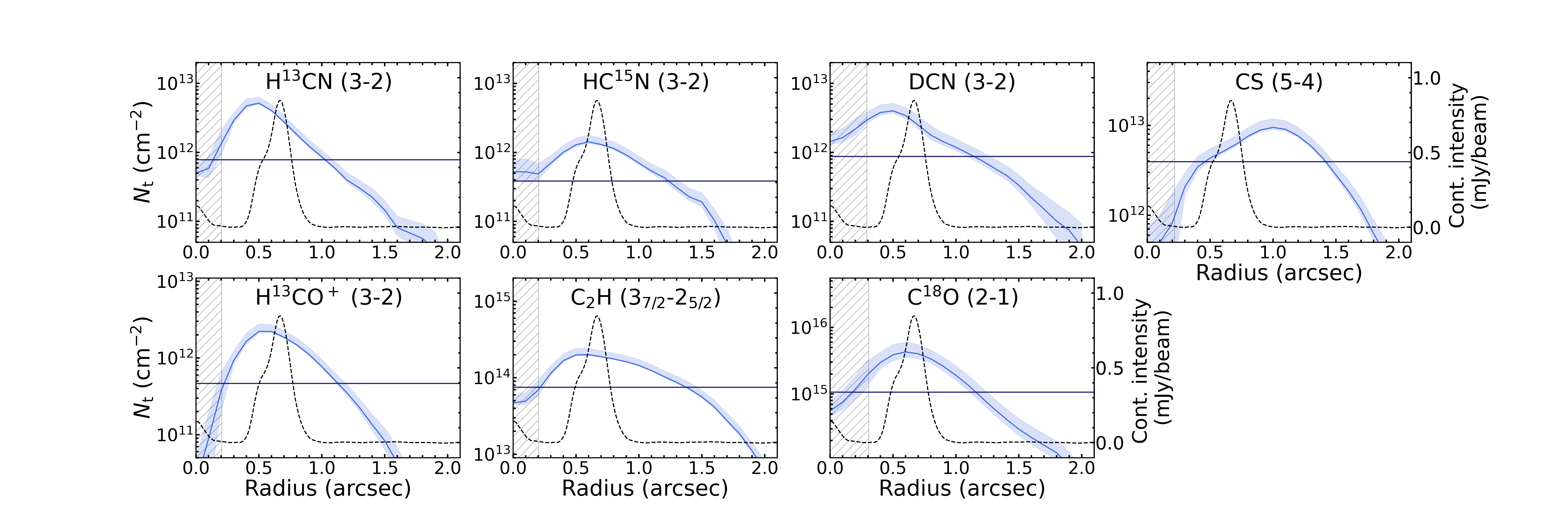}
    \caption{Radial profiles of the column density for the observed molecules. The solid blue lines refer to the value extracted for an excitation temperature of $T_\mathrm{ex}$=30\,K, while the ribbons show the uncertainty, which takes into account the propagated uncertainty on the flux density and the systematic uncertainty on the excitation temperature, which has been varied between 20\,K and 50\,K \citep{bib:facchini2021chemical}. The 10\% absolute flux calibration uncertainty is not included. The horizontal dark blue line shows the disk-averaged value, listed in Table~\ref{tab:line_par}. The dashed black line shows the azimuthally averaged radial profile of the 855$\mu$m continuum emission \citep{isella2019detection}. Hatched regions indicate the beam major axis of the cubes.}
    \label{fig:N}
\end{figure*}

From the radial profiles of the integrated intensity (Fig.~\ref{fig:profiles}), we extracted the molecular column density as a function of radius for all the molecules listed in Table~\ref{tab:line_par} (see Fig.~\ref{fig:N}), by computing the column density along $0\farcs1$ wide annuli, using the line parameters listed in Table~\ref{tab:line_par}. We varied the excitation temperatures from 20\,K to 50\,K, as for the disk-averaged column density. The solid blue lines in Fig.~\ref{fig:N} show the radial profiles of the total column density obtained for the different molecules, at the assumed $T_\mathrm{ex}=30$\,K.

For the \ce{H2CO} molecule we also applied a radial rotational diagram, to retrieve both $T_\mathrm{ex}(r)$ and $N_\mathrm{t}(r)$, where $r$ is the cylindrical radius. We divided the disk into $0\farcs1$ wide annuli and we sampled the posterior distribution for each annulus using an MCMC method with the \texttt{emcee} package \citep{foreman2013emcee}, with 128 walkers, 5000 burn-in steps and 500 steps. We considered a uniform prior with $T_\mathrm{ex}\in(0,100)$\,K, and $N_\mathrm{t}\in(10^{11}, 10^{16})$\,cm$^{-2}$. Radially resolved \ce{H2CO} excitation temperature and column density are shown in the bottom panels in Fig.~\ref{fig:rot_diag}. Ribbons indicate the 16th and 84th  percentiles of the posterior distribution, and the propagated uncertainty on the integrated intensity has been considered when generating the likelihood. The column density peaks at smaller radii with respect to the submillimeter continuum bright ring, which is consistent with the radial profile of the integrated intensity for the two \ce{H2CO} lines, in Fig.~\ref{fig:profiles}.
The radial profile of the excitation temperature is almost flat, with $T_\mathrm{ex}$ between 20-30~K already from $r=0\farcs4$ inside the dust cavity (discussed in Sect.~\ref{subsec:h2co_origin}). The higher excitation temperature at smaller radii is likely due to the dominant contribution from the inner emission observed in the faint and higher energetic \ce{H2CO} (3$_{2,1}$-2$_{2,0}$) line, showing a hot spot at the cavity wall near PDS~70b (see Sect.~\ref{subsec:faint_lines_H2CO} for deeper discussion).

\section{Discussion} \label{sec:discussion}

\subsection{Molecular emission peaking at the cavity wall}\label{subsec:discuss_lines}

From the integrated intensity maps in Fig.~\ref{fig:M0} and their azimuthally averaged radial profiles in Fig.~\ref{fig:profiles}, we found that the emission morphologies in high resolution data are in good agreement with \cite{bib:facchini2021chemical}. 

\begin{figure}[ht!]
    \centering
    \includegraphics[width=0.49\textwidth]{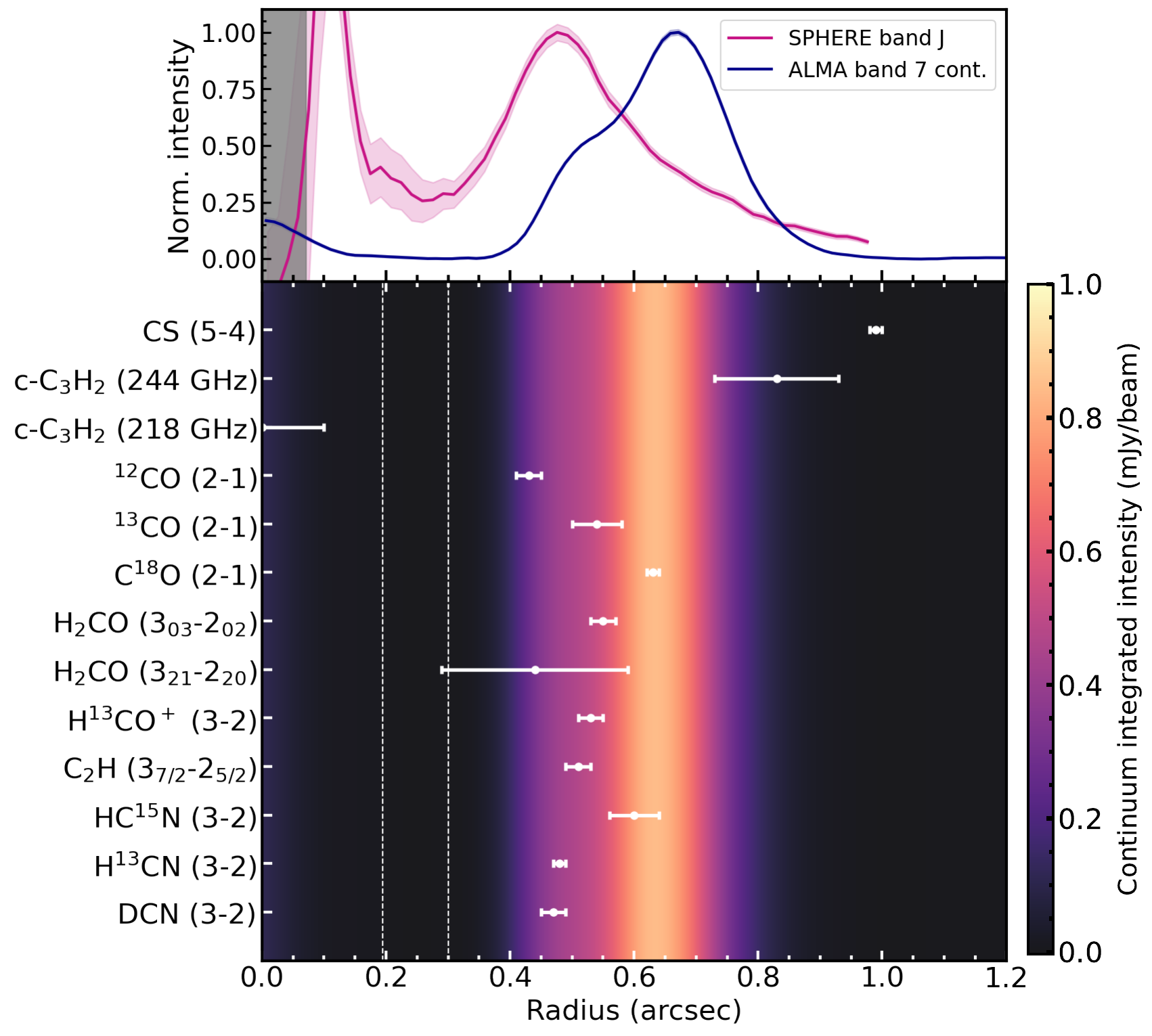}
    \caption{Radial location of the peak of the emission of the various molecular tracers, compared to radial profiles of SPHERE scattered light and ALMA band 7 continuum intensity. Top panel: Radial profiles of the ALMA band 7 continuum intensity \citep[blue line,][]{isella2019detection} and SPHERE $J$-band polarized intensity \citep[pink line,][]{keppler2018discovery}. The gray shadow shows the radius of the coronagraph in the $J$-band observations \citep{keppler2018discovery}. Bottom panel: Radial position of the peak of the emission and its related uncertainty, for each detected line. The background color-bar refers to the radial profile of the ALMA band 7 continuum intensity. The two vertical dashed white lines indicate the planets position \citep{wang2021constraining}.}
    \label{fig:peak_cont}
\end{figure}

To highlight the main emitting features, we list the radial location of the emission peak for each molecule in Table~\ref{tab:radii}, and we compare them in Fig.~\ref{fig:peak_cont} (bottom panel), together with the radial profile of the ALMA band 7 continuum intensity (top panel, blue line, \citealt{isella2019detection}) and of the SPHERE $J$-band polarized intensity (top panel, pink line, \citealt{keppler2018discovery}). We extracted the radial position of the peaks by fitting a multiple-Gaussian profile (triple for \ce{^12CO} and \ce{H2CO}, double for the other molecules) to each radial profile in Fig.~\ref{fig:profiles}, using \texttt{scipy.optimize.curve\_fit} (see Fig.~\ref{fig:profiles_fit}). Uncertainties on the peaks position have been obtained from the \texttt{scipy} fit, taking into account also the uncertainties on the integrated intensity. For the faint \ce{c-C3H2} transitions we could not perform a Gaussian fitting, and we therefore provide the peak position along with 1/4 of the beam major axis as an associated uncertainty.
In Table~\ref{tab:radii} we also list the radius corresponding to the 68\% and 90\% of the radius enclosing the whole line flux, as computed from a cumulative flux radial profile \citep{facchini2019high, sanchis2021measuring}.

As apparent from the bottom panel in Fig.~\ref{fig:peak_cont}, molecular emission does not necessarily correlate with the dust distribution. This is also consistent with what was shown by the MAPS Large Program \citep{oberg2021molecules} through high resolution line emission observations of five protoplanetary disks: all disks show a variety of chemical substructures which can critically differ from the dust ones \citep{law2021molecules, law2021emission}.

For the specific case of the PDS~70 disk, high resolution line emission observations show that most of the molecules present a peak of the emission at smaller radii with respect to the submillimeter continuum bright ring (bottom panel of Fig.~\ref{fig:peak_cont}).
Different tracers are affected by specific chemical and physical conditions, and thus peak at different radii, not always tracing the underlying gas density structure.


The emission of CO isotopologs is mainly affected by their different optical depth, with the optically thicker $^{12}$CO tracing the gas temperature, and thus peaks at an inner radius ($\sim0\farcs43$) with respect to the submillimeter continuum bright ring ($\sim0\farcs67$), while the optically thinner C$^{18}$O better traces the bulk of the gas mass and peaks at $\sim0\farcs63$ \citep{portilla2023constraining}. This trend in the peak of CO isotopologs is also consistent with planet-disk interaction models \citep{pinilla2012ring, facchini2018inferring} and with what it has been found for other disks \citep[e.g., ][]{perez2014co, van2015resolved, leemker2022gas}.

Except for the CO isotopologs, another effect which can play a role in shaping the emission of the other molecular tracers is the irradiation field. As the two forming planets carved a large dust cavity, radiation from the central star can penetrate freely and efficiently irradiate the cavity wall. PDS~70 is the first transition disk with clear evidence of a prominent cavity wall, with a steep rise in emitting height and a vertical dip in the \ce{^12CO} (3-2) line \citep{law2024mapping}.  The emission of molecules whose production is enhanced in a strong irradiation field is expected to peak at the warm, strongly irradiated cavity wall rather than following the gas density distribution \citep{portilla2023constraining}. Among these tracers, HCN isotopologs, \ce{C2H}, and \ce{H^13CO+} peak inside the bright submillimeter ring, showing a correlation with the peak in the scattered light profile (see Fig.~\ref{fig:peak_cont}) which traces the rise of the density of small grains from the cavity wall \citep{keppler2018discovery, portilla2023constraining}. In particular, \ce{C2H} is expected to be enhanced in a strong UV-field, and also HCN, as it is produced by UV-pumped H$_2$. Similarly, ions such as H$^{13}$CO$^+$ are formed in a strong X-ray field environment outside the water snowline. While the submillimeter continuum profile traces the distribution of large grains at the midplane, the NIR scattered light profile traces the disk illumination pattern. The correlation between the peak of the emission for the molecules listed above and the NIR peak is therefore consistent with their production being favored by photons from the star which can freely reach the cavity wall.

HCN isotopologs also show different emission morphologies, with H$^{13}$CN being the most compact one, while HC$^{15}$N and DCN present a more extended emission, possibly linked to active in-situ fractionation pathways. In particular, isotope selective photodissociation and the effective self-shielding of \ce{N2} could strongly affect \ce{H^13CN} and \ce{HC^15N} inside the UV-irradiated cavity \citep[][]{heays2014isotope, visser2018nitrogen, hily2019multiple}. In particular, this process could explain the observed shift in the \ce{H^13CN} and \ce{HC^15N} peaks: while \ce{H^13CN} peaks at an inner radius, \ce{HC^15N} peaks at an outer radius (similar to \ce{C^18O}), which could reflect the fact that \ce{N2} easily self-shields at a higher gas density, with respect to \ce{N^15N}.  On the other hand, low temperature isotope exchange reactions may play a crucial role in producing the outer emission shoulders in the \ce{HC^15N} and DCN emission \citep{millar1989deuterium, roueff2015isotopic, huang2017alma, oberg2020tw, cataldi2021molecules, romero2023cold}. Fractionation processes presented above result in different HCN isotopologs peaking at different radii: $\sim0\farcs48$ for \ce{H^13CN}, $\sim0\farcs60$ for \ce{HC^15N}, and $\sim0\farcs47$ for DCN. 
A deeper characterization of the fractionation levels of the HCN molecule in the PDS~70 disk, their radial profiles and implications for the planet-forming environment will be investigated in a future work, as well as further modeling to constrain the role of the strongly irradiated cavity wall in setting the observed emission morphology.

In this picture, the bright ring in the submillimeter continuum emission appears to induce a chemical separation, between an inner warm and strongly irradiated cavity wall and an outer low-temperature and low-density environment. While most of the molecules show an inner bright ring, a few tracers also present outer peaks or shoulders of emission (see Fig.~\ref{fig:profiles}), which are due to different chemical reactions specifically unlocked by environmental conditions set beyond the continuum bright ring, as we previously outlined for the case of the DCN molecule. Similarly, also \ce{H2CO} shows an outer peak at $\sim$1.27$\as$, which we discuss further in Sect.~\ref{subsec:h2co_origin}. 

In this context, we note that all C-rich molecules present bright emission beyond the submillimeter dust ring: CS peaks at $\sim1\as$, \ce{c-C3H2} (244 GHz) peaks at $\sim0\farcs83$, and \ce{C2H} shows a bright outer shoulder beyond $\sim0\farcs87$. This could be indicative of a carbon-rich chemistry boosted in the outer disk, and a corresponding radially increasing gas-phase C/O ratio, which we discuss further in Sect.~\ref{subsec:C/O}. \ce{c-C3H2} (218 GHz) is the only exception, with centrally peaked compact emission: this is consistent with the higher upper state energy of the transition ($E_\mathrm{u}=61$~K), and thus better traces high temperature at smaller radii.

In Fig.~\ref{fig:peak_cont_peak} we compare the position of the peaks of the emission extracted from the integrated intensity (in white) and from the peak intensity (in red) radial profiles, obtained by multiple-Gaussian fitting (see Fig.~\ref{fig:peak_profiles}). 
Radial profiles of the peak intensity were obtained by azimuthally averaging the peak intensity maps, which show the peak of the spectrum for each spatial pixel (see Fig.~\ref{fig:M8}).
Peak intensity profiles better trace optical depth, as they do not depend on the line width like integrated intensity profiles, which include contribution at the same location both from the line center and line wings, where optical depth can change \citep{weaver2018empirical, rosotti2021high}. As line width increases at smaller radii, we expect the peaks from integrated intensity profiles to be at an inner radius with respect to peaks from peak intensity profiles: this is visible in Fig.~\ref{fig:peak_cont_peak}. However, the peak positions are consistent within the error bars for most of the molecules, and thus robustly confirm the observed trend of the peaks of line emission, which are inside the millimeter dust peak for most of the molecular tracers. This result is also consistent with the fact that the peaks in Fig.~\ref{fig:M8} are spatially and spectrally resolved: beam dilution, which is expected to affect the peak position extracted from peak intensity profiles \citep{leemker2022gas}, is thus negligible in this case.

\begin{table}[h!]
  \renewcommand{\arraystretch}{1.3}
  \caption{
  Radial position of the peak of the integrated intensity profiles, and radii enclosing 68\% and 90\% of the total flux.
  }
  \label{tab:radii}
  \begin{threeparttable}
  \begin{tabular}{c c c c }
      \hline
      \hline
      Transition\tnote{a} & $R_\mathrm{peak}$ \tnote{b} [$\as$] & $R_{68}$ [$\as$] & $R_{90}$ [$\as$] \\
      \hline
      $^{12}$CO (2-1) & 0.43 $\pm$ 0.02 & 1.38 & 1.81 \\
      $^{13}$CO (2-1) & 0.54 $\pm$ 0.04 & 1.11 & 1.49 \\
      C$^{18}$O (2-1) & 0.63 $\pm$ 0.01 & 0.96 & 1.34 \\
      H$_2$CO (3$_{0,3}$-2$_{0,2}$) & 0.55 $\pm$ 0.02 & 1.27 & 1.70 \\
      H$_2$CO (3$_{2,1}$-2$_{2,0}$) & 0.44 $\pm$ 0.15 & - & - \\
      C$_2$H ($3_{7/2}$-$2_{5/2}$) & 0.51 $\pm$ 0.02 & 1.19 & 1.60 \\
      H$^{13}$CN (3-2) & 0.48 $\pm$ 0.01 & 0.78 & 1.14 \\
      HC$^{15}$N (3-2) & 0.60 $\pm$ 0.04 & 0.99 & 1.30 \\
      DCN (3-2) & 0.47 $\pm$ 0.02 & 0.99 & 1.43 \\
      H$^{13}$CO$^+$ (3-2) & 0.53 $\pm$ 0.02 & 0.85 & 1.15 \\
      CS (5-4) & 0.99 $\pm$ 0.01 & 1.22 & 1.55 \\
      c-C$_3$H$_2$ (244 GHz) & 0.83 $\pm$ 0.1\tnote{c} & - & - \\
      c-C$_3$H$_2$ (218 GHz) & 0.0 $\pm$ 0.1\tnote{c} & - & - \\
      Continuum 855$\mu$m & 0.646 $\pm$ 0.002 & 0.71 & 0.82 \\
      \hline
      \hline
  \end{tabular}
  \begin{tablenotes}
      \item[a] See table \ref{tab:imaging_molecules} for quantum number formatting.
      \item[b] Radial position of peaks were obtained from multiple-Gaussian fitting, except for the faint \ce{c-C3H2} transitions.
      \item[c] $R_\mathrm{peak}$ refers to the peak of the integrated intensity profile, while the uncertainty is assumed to be a quarter of the beam major axis, as Gaussian fitting was not possible.
  \end{tablenotes}
  \end{threeparttable}

\end{table} 
  
We summarize the peak positions extracted from peak intensity profiles in Table~\ref{tab:radii_peaks}: the two only tracers with inconsistent peaks are \ce{C2H} and \ce{^13CO}, with the peak position from peak intensity profiles being further out with respect to integrated intensity profiles, as expected. The most critical case is \ce{C2H} ($R_\mathrm{peak} = 0\farcs51$ from integrated intensity, $R_\mathrm{peak} = 0\farcs83$ from peak intensity): this can be due to the fact that the molecule presents two similar peaks at two different radii (see Fig.~\ref{fig:M8}), which weigh differently in the peak intensity and in the integrated intensity profiles, as explained above. Moreover, from the radial profile of the peak intensity in Fig.~\ref{fig:peak_profiles}, we notice that the \ce{H2CO} molecule is the only one showing a gap beyond the submillimeter bright ring, followed by an outer ring of emission (see also Sect.~\ref{subsec:faint_lines_SO}).

\subsection{Comparison with other transition disks}\label{subsec:discuss_comparison}

Substructures in the line emission of molecular tracers are the product of a variety of physical and chemical pathways, and are therefore not always colocated with dust substructures \citep[][and references therein]{oberg2021molecules}. We showed that the PDS~70 system fully agrees with this scenario (see Sect.~\ref{subsec:discuss_lines}), but in this case the efficient irradiation of the cavity wall could play an important role in enhancing molecular emission of specific tracers inside the dust cavity, which could be a specificity of transition disks.

In this context, we compare our result with two other transition disks around HD~169142 \citep{garg2022kinematic, booth2023tracing} and HD~142527 \citep{temmink2023investigating}, for which a chemical overview has been provided through ALMA line emission observations. As different disks around different stars present different temperature structure, irradiation field, size and mass, in order to provide a consistent comparison and look for a possible correlation between molecular emission and the cavity wall, we extracted the ratio between the radius corresponding to the peak of the emission of various molecular tracers and the one of the innermost bright ring in the continuum emission (we call it “peak ratio” for simplicity).  

CO isotopologs and CS show the most consistent trend among the three disks. CS always peaks outside the innermost continuum bright ring, with a peak ratio of $\sim$1.5, 1.5 and 2.2 for the PDS~70, HD~142527, and HD~169142 disk respectively. In this context, \cite{temmink2023investigating} extracted a CS/SO ratio >1, as we obtained for the PDS~70 disk in this work (see Sect.~\ref{subsec:C/O}), while \cite{booth2023tracing} presented a radially increasing CS/SO ratio. These first attempts highlight the importance of better constraining the C/O ratio and its radial variations across transition disks directly from observations. The volatile C/O ratio has been shown to be strongly linked to planetary accretion \citep{jiang2023chemical}: on one hand, the C/O ratio of the gas accreted by the planet will be inherited by the planetary atmosphere, but on the other hand, a forming planet that is carving a gap in the disk is also expected to locally change the C/O ratio in the gas phase due to, for example, the sublimation of C-rich molecules.

The consistent trend shown by peaks of CO isotopologs among the three disks robustly confirm the dominant effect of their different optical depths, as we previously discussed in Sect.~\ref{subsec:discuss_lines}. In particular, the peak ratios increase with decreasing optical depths: for \ce{^12CO} it is $\sim$0.7, 0, and 0.6, for \ce{^13CO} is $\sim$0.8, 0.7 and 0.8, and for \ce{C^18O} is $\sim$0.9, 1.0, and 1.0, for PDS~70, HD~142527, and HD~169142, respectively. In this context, a critical difference between the three sources is the presence of a gas gap/cavity: while both the PDS~70 and the HD~169142 disk present a gas gap/cavity clearly visible in the \ce{^12CO} emission, \ce{^12CO} extends towards the disk center for the HD~142527 disk.

On the other hand, except for CO isotopologs and CS, there are not any other consistent trends among the peak ratios for other molecular tracers. For example, for the disk around HD~169142 \citep{booth2023tracing}, both \ce{H2CO} and \ce{c-C3H2} peak further out (peak ratios of $\sim$5 and 4 respectively), with respect to the PDS~70 disk (peak ratios of $\sim$0.8 and 1.2 respectively). Similarly, if we compare peak ratios of observed lines of HCN isotopologs, we obtain $\sim$0.74 for \ce{H^13CN} in the PDS~70 disk and $\sim$1.5 for \ce{HCN} in the HD~142527 disk. Even if these are different isotopologs and could therefore be sensitive to different chemical pathways, we would have anyway expected a peak of the emission inside the inner cavity, if the irradiation of the cavity wall had played a major role. However, \cite{cazzoletti2018cn} implemented a set of disk models through the thermo-chemical code DALI \citep{bruderer2012dali, bruderer2013dali}, to show that rings in the CN emission can be traced back to its main formation pathway through UV-pumped \ce{H2}, which is mostly abundant in a ring-shaped region close to the surface layer of the disk. This process does not require the presence of a cavity, but results in a CN ring also for smooth disks. In particular, CN rings are at larger distances from the central star in disks around Herbig stars than T~Tauri stars, due to their different UV fields. HCN is also produced by UV-pumped \ce{H2}, and we therefore expect it to be strongly affected by different UV fields, which could partially explain the observed difference between the radial position of the HCN isotopologs rings, as HD~169142 and HD~142527 are Herbig stars, while PDS~70 is a T~Tauri. Moreover, PDS~70 shows the largest gas cavity in the \ce{^12CO} emission ($\sim$43~au with respect to $\sim$15~au for the HD~169142 disk). PDS~70 is the only disk for which a vertical dip at the cavity wall is clearly visible in \ce{^12CO} line emission, revealing a prominent, warm, and strongly irradiated cavity wall \citep{law2024mapping}.

Nevertheless, the observed diversity in the molecular emission morphology throughout the three transition disks suggests that a further and deeper characterization of line emission observations and further model calculations in transition disks is needed in order to shed light on the effect of the inner cavity on the disk chemistry, and to assess if the observed chemical properties of the planet-hosting disk around PDS~70 make it a peculiar planet-forming environment.

\subsection{Azimuthally asymmetric emission of \texorpdfstring{$\,H_2CO$}{H2CO}}\label{subsec:faint_lines_H2CO}

\begin{figure*}[ht!]
    \centering
    \includegraphics[scale=0.23]{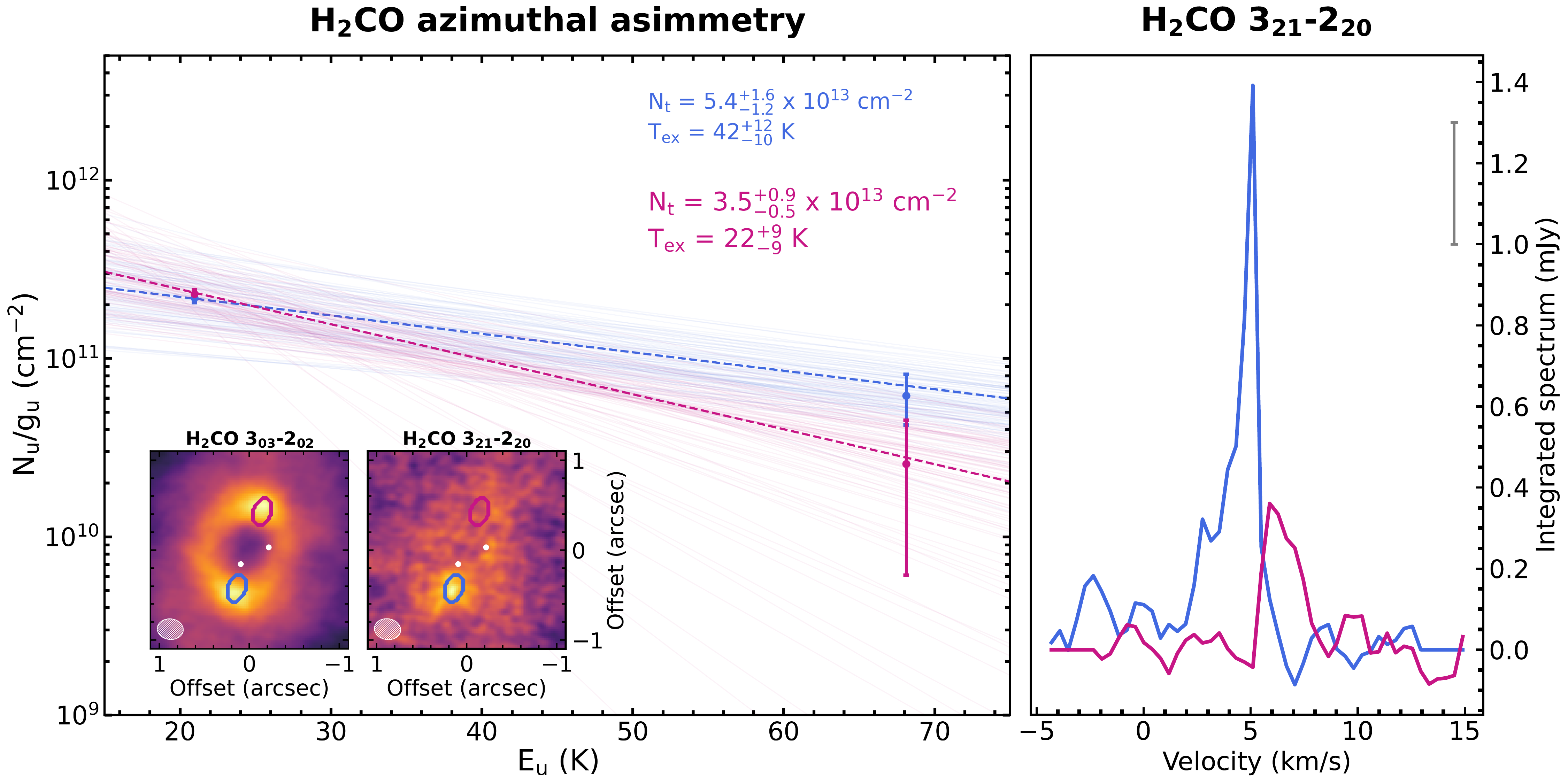}
    \caption{Azimuthal asymmetry of the H$_2$CO 3$_{2,1}$-2$_{2,0}$ line emission. Left panel: Rotational diagram analysis applied to the two regions, highlighted in the integrated intensity maps of the two lines of the H$_2$CO molecule: the pink region corresponds to PA=0$^\circ$, while the blue one is at PA=180$^\circ$ along the major axis. The dashed pink and blue lines correspond to the best fit $N_\mathrm{t}$ and $T_\mathrm{ex}$ (reported in the top right corner) obtained for the pink and blue region, respectively. Right panel: Spatially integrated spectra of the H$_2$CO 3$_{2,1}$-2$_{2,0}$ line in the pink region and the blue region (lines in corresponding colors).
    The gray error bar in the top right of the panel is the uncertainty on the spatially integrated flux derived from the non-JvM corrected cube, which is the same for each velocity bin.}
    \label{fig:h2co_asymmetry}
\end{figure*}

The two forming planets in the cavity of the PDS~70 disk have been shown to be actively accreting material from their environment, through ultraviolet (for PDS~70b) and H$\alpha$ line emission observations, which have been used to derive an estimate of the mass accretion rate over PDS~70b and c \citep{wagner2018magellan, aoyama2019constraining, haffert2019two, zhou2021hubble}. Similarly, we could expect that the process of accretion onto the forming planets results in enhanced emission of particular molecules observed with ALMA \citep{cleeves2015indirect}. In this context, we looked for accretion signatures in the PDS~70 disk through high excitation lines, as they could be more sensitive to local heating mechanisms. We found an azimuthal asymmetry in the H$_2$CO (3$_{2,1}$-2$_{2,0}$) faint line emission, with a localized hot spot at the cavity wall at the same azimuth as PDS~70b, which could be linked to planetary accretion heating. This is particularly evident in the peak intensity map in Fig.~\ref{fig:M8}, but also in the integrated intensity map in Fig.~\ref{fig:M0}. In order to test the significance of the azimuthal asymmetry, we extracted the spatially integrated spectrum within an area of the beam size localized on the emission feature, at an angular position of 0$^\circ$ along the major axis, and from the opposite angular position at 180$^\circ$, as we show in the integrated intensity maps in the left panel of Fig.~\ref{fig:h2co_asymmetry}. The integrated spectra are shown in the right panel of Fig.~\ref{fig:h2co_asymmetry}, where the solid blue line refers to PA=0$^\circ$, while the pink one to PA=180$^\circ$. These do show an asymmetry between the blue and red-shifted parts of the disk. Nevertheless, the line showing the azimuthal asymmetry is faint, and the statistical significance of the spectral difference is low (2.5$\sigma$). We extracted the significance by taking the ratio of the difference between the two peaks and $\sqrt{2}\,\times$ the spectrum uncertainty, which has been evaluated as the standard deviation of the fluxes from 50 random beam-sized regions outside the disk emitting area, in a random velocity channel. The uncertainty was extracted from the non-JvM corrected cube, since the JvM correction underestimates the uncertainty on the flux density of point sources by its correction factor, as we highlighted in Sect.~\ref{sec:data}.

To further explore the physical significance of the asymmetry, we applied the rotational diagram analysis in the two regions (pink and blue, as we previously described) to the two H$_2$CO (3$_{0,3}$-2$_{0,2}$) and (3$_{2,1}$-2$_{2,0}$) lines, following the procedure we explained in Sect.~\ref{subsec:N_av}. The result is presented in the left panel of Fig.~\ref{fig:h2co_asymmetry}, where the pink line refers to the results for the pink region, while the blue one refers to the blue region. As expected, we obtained a higher excitation temperature ($\sim$42\,K) at the peak position, with respect to the opposite one ($\sim$22\,K), but the two temperatures are consistent within the error bars, making the difference not statistically significant. We notice that we did not detect significant asymmetries in the H$_2$CO (3$_{0,3}$-2$_{0,2}$) emission, or in any other accretion tracers. However, this result highlights the importance of further exploring planetary signatures in high energy transitions, as the only feature we identified is from the higher energy H$_2$CO line.

\subsection{On the origin of the observed \texorpdfstring{$\,H_2CO$}{H2CO} emission}\label{subsec:h2co_origin}

As we previously highlighted in Sect.~\ref{subsec:m0} and \ref{subsec:N_rad}, the emission of the bright \ce{H2CO} (3$_{0,3}$-2$_{0,2}$) line presents a peak at smaller radii with respect to the submillimeter continuum ring (see Fig.~\ref{fig:profiles} and the bottom-left panel of Fig.~\ref{fig:rot_diag}), and an outer shoulder of emission extending beyond the outer edge of the millimeter continuum disk. The presence of an outer shoulder of emission for the H$_2$CO molecule has been observed in other sources \citep{loomis2015distribution, carney2017increased, guzman2021molecules, van2021tw, booth2023tracing, temmink2023investigating}, but its interpretation is not straightforward, as it has been linked both to a possible gas-phase or solid-state origin. In particular, the dominant gas phase formation pathway is \citep{fockenberg2002temperature, atkinson2006atmospheric}:

\begin{equation}
    \mathrm{CH}_3 + \mathrm{O} \rightarrow  \mathrm{H}_2\mathrm{CO} + \mathrm{H},
    \label{eq:h2co_gasphase}
\end{equation}
while the main grain surface formation route is through hydrogenation of CO ice (\citealt{Hiraoka1994formation}; \citeyear{hiraoka2002formation}; \citealt{watanabe2002efficient}; \citeyear{watanabe2004hydrogenation}; \citealt{hidaka2004conversion}; \citealt{fuchs2009hydrogenation}):
\begin{equation}
    \mathrm{CO} \xrightarrow{\text{H}}  \mathrm{HCO} \xrightarrow{\text{H}} \mathrm{H}_2\mathrm{CO}.
    \label{eq:h2co_grain}
\end{equation}

\begin{figure}[ht!]
    \includegraphics[scale=0.21]{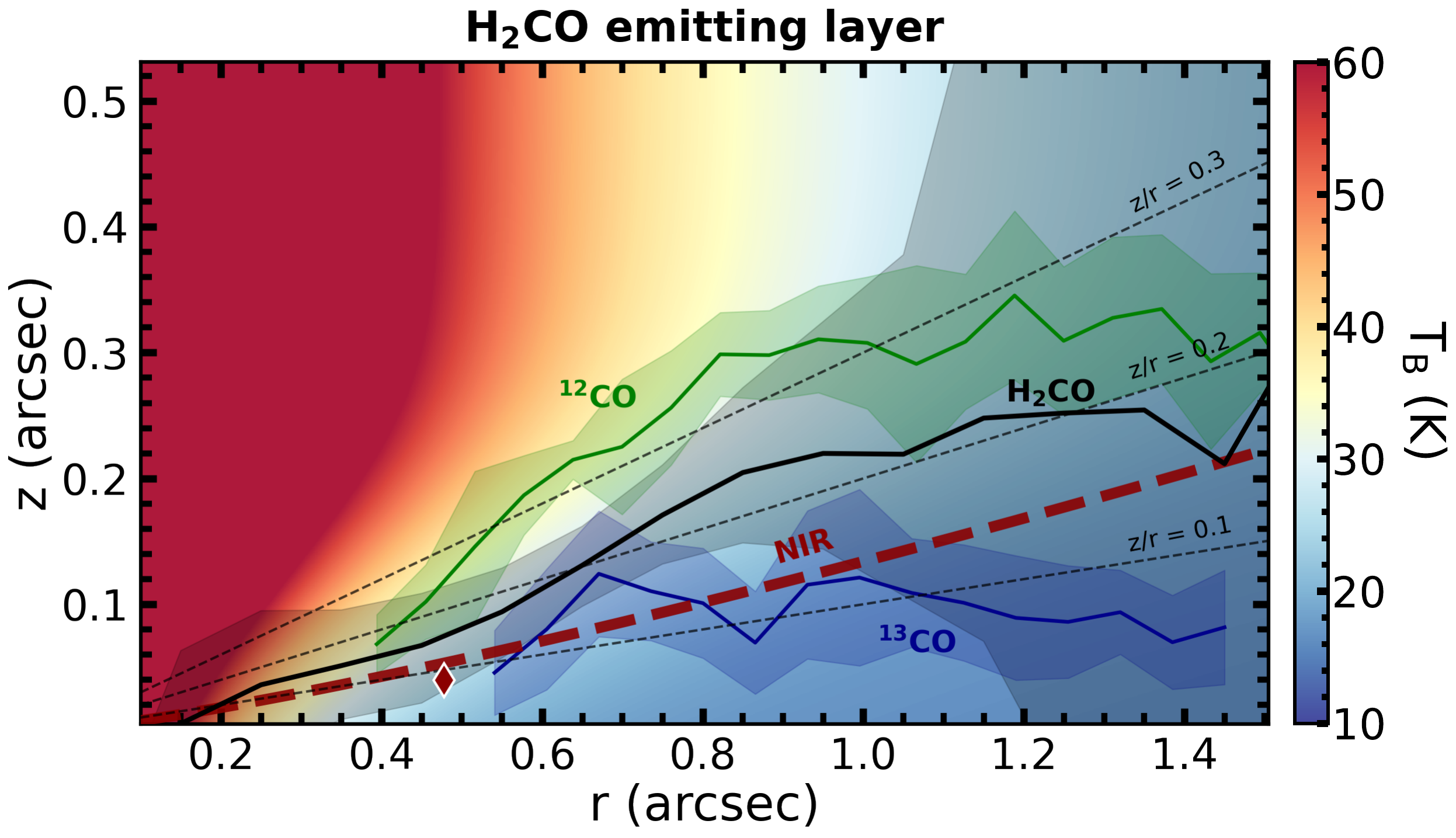}
    \caption{\ce{H2CO} emitting layer height as a function of radius (solid black line), and the related uncertainty (black ribbons), compared to the 2D temperature structure in the background \citep{law2024mapping}.
    The two solid green and blue lines refer to the \ce{^12CO} and \ce{^13CO} emission surfaces, with related uncertainties, respectively \citep{law2024mapping}.  The dashed red line refers to the NIR scale height as a function of radius, from \cite{keppler2018discovery}, while the red diamond is the height of the NIR emitting layer at the NIR ring $\sim0\farcs48$ \citep{law2024mapping}. Dashed black lines indicate fixed $z/r$ layers.}
    \label{fig:h2co_emitting_layer}
\end{figure}

If \ce{H2CO} originates from ice sublimation we would expect some correlation between \ce{H2CO} emission and dust grain locations, either through a correlation with the millimeter continuum bright ring (desorption from millimeter dust), or with the \ce{H2CO} emitting layer being consistent with or below the near infrared surface, where the $\mu$m dust is expected to reside. We therefore retrieved the \ce{H2CO} emitting layer, from the 2D temperature structure of the PDS~70 disk, which was extracted by \cite{law2024mapping}, combining CO isotopologs and \ce{HCO^+}. The rotational diagram assumes that the excitation temperature $T_\mathrm{ex}(r)$ is equal to the kinetic temperature $T(r,z)$ (LTE). We therefore extracted the emitting layer $z(r)$ by inverting the $T(r,z)$ relation \citep{law2024mapping}, as it has been previously performed by \cite{ilee2021molecules} for the case of complex organic molecules in the MAPS sample. We obtained that the H$_2$CO emission originates from $z/r\sim 0.2$ (see Fig.~\ref{fig:h2co_emitting_layer}, solid black line), under the assumption of a geometrically thin emitting layer. We highlight that the LTE approximation is valid at the inferred layer, as the critical density of $\sim6\times 10^5$cm$^{-3}$ for \ce{H2CO} at the derived excitation temperatures is reached at $z/r\gtrsim0.3$. The inferred layer for \ce{H2CO} is overlaid on the 2D temperature structure \citep{law2024mapping} and compared to the emission surfaces of \ce{^12CO} and \ce{^13CO} (solid green and blue lines).

The trend of the \ce{H2CO} excitation temperature is almost flat, and between 20\,K and 30\,K, which suggests a low-temperature, radical-radical gas-phase formation pathway. This scenario is also supported by the absence of an evident correlation between the H$_2$CO emission and the submillimeter continuum emission, and the detection of molecular emission from various molecules such as C$_2$H, \ce{c-C3H2}, H$^{13}$CN whose gas-phase production is boosted in UV-exposed regions. Moreover, thermal-desorption from icy grains is an unlikely process to release \ce{H2CO}, as both the H$_2$CO excitation temperature and the $^{12}$CO kinetic temperature are lower than the typical H$_2$CO sublimation temperature of $\sim$70\,K \citep{noble2012desorption, fedoseev2015experimental, van2020temperature}, for $r\gtrsim0\farcs4$. 

Nevertheless, the grain-surface formation pathway cannot be completely ruled out, as H$_2$CO could be released also through nonthermal desorption, such as photo-desorption, in UV-exposed regions. This scenario is consistent with the low excitation temperature we inferred ($\sim$20-30\,K), corresponding to CO freeze-out, and with the detection of bright UV-boosted molecules. Moreover, this formation route is supported by the fact that the radial profile of the \ce{H2CO} column density (see right panel, bottom row in Fig. \ref{fig:rot_diag}) peaks at the position of the CO snowline \citep{oberg2017h2co, guzman2021molecules}, which lays between $\sim0\farcs5-0\farcs75$ \citep{law2024mapping}. In this context, desorption from $\mu$m grains cannot be ruled out, as the inferred \ce{H2CO} emitting layer is slightly above the NIR surface (dashed red line in Fig.~\ref{fig:h2co_emitting_layer}, \citealt{keppler2018discovery}), but uncertainties do not allow one to distinguish the two scenarios.
However, it is important to note that the efficiency of nonthermal desorption from icy grains is not well constrained \citep{oberg2017h2co}, and distinguishing between grain-surface and gas-phase pathways is not trivial, as one can continue to contribute also at the onset of the other \citep{walsh2014complex, loomis2015distribution, oberg2017h2co}. 

\begin{figure*}[ht!]
    \centering
    \includegraphics[scale=0.55]{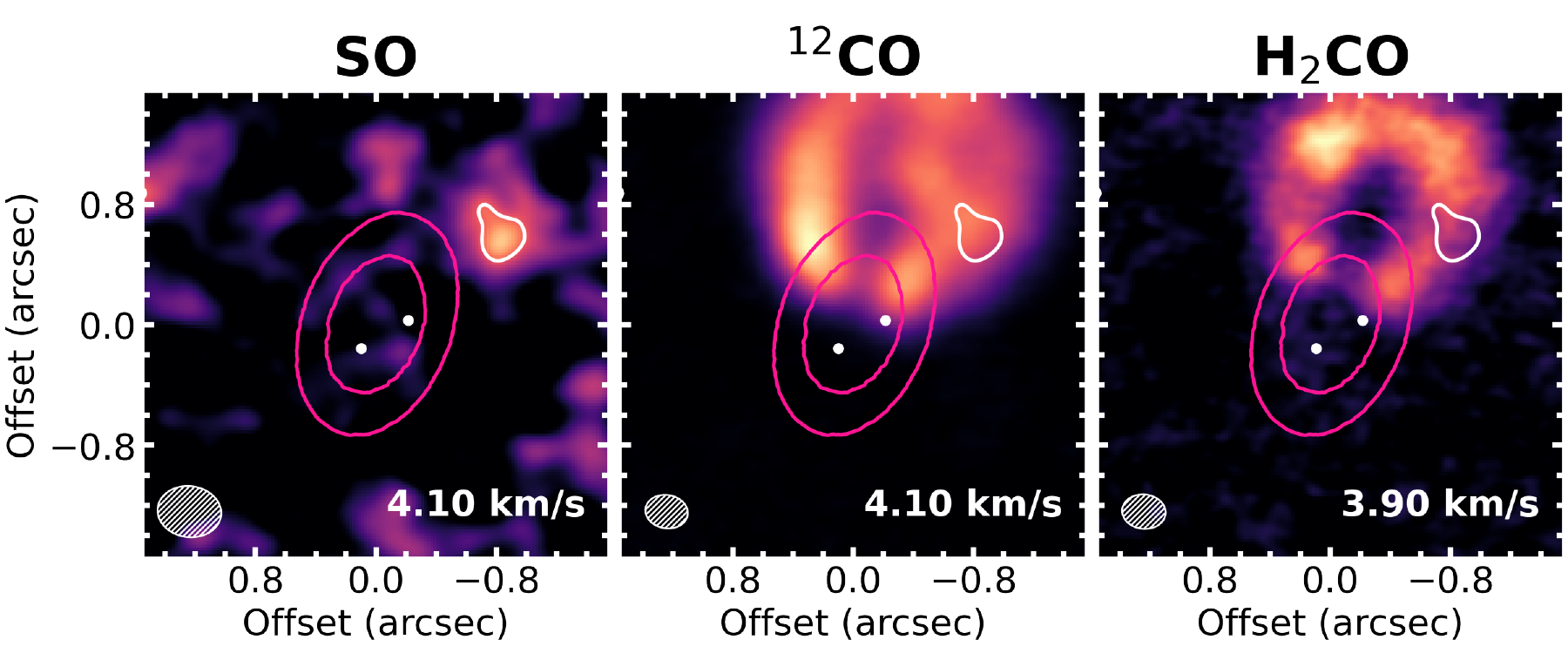}
    \caption{Channel maps around 4~km~s$^{-1}$ showing the spatial distribution of the emission of three different molecules: SO (stacked), $^{12}$CO and H$_2$CO, from left to right. The white contour shows the 3$\sigma$ level of the SO emission, which is reported in the other channel maps for reference. The pink contours indicate the bright ring in the band 7 continuum emission \citep{isella2019detection}, while the two white dots mark the position of the two protoplanets \citep{wang2021constraining}. The white ellipse in the bottom left side of each panel is the beam of each imaged cube.
}
    \label{fig:SO_signature}
\end{figure*}

Interestingly, the \ce{H2CO} excitation temperature we obtained (between 20\,K and 30\,K) is lower than the \ce{H2CO} freeze out temperature of $\sim70$\,K \citep{noble2012desorption, fedoseev2015experimental, van2020temperature}, as it has been also shown for other sources \citep[][]{van2021tw, temmink2023investigating}: this implies that \ce{H2CO} needs to be efficiently produced, either through gas-phase reactions or nonthermally released from icy grains, with faster timescales with respect to the freeze-out one.

\subsection{Low emitting layer of c-\texorpdfstring{$\,C_3H_2$}{C3H2}}\label{subsec:faint_lines_C3H2}

\ce{c-C3H2} has been previously detected in a few other disks \citep[][]{bergin2016hydrocarbon, ilee2021molecules, booth2023tracing}. In particular, \cite{ilee2021molecules} presented the detection of \ce{c-C3H2} in four disks in the MAPS sample (GM Aur, AS 209, HD 163296, and MWC 480). They extracted a disk average column density of \ce{c-C3H2} between 1.1$\times$10$^{12}$ cm$^{− 2}$ and 7.0$\times$10$^{12}$ cm$^{− 2}$ across the sample, which is consistent with the disk average column density of $\sim$6.8$\times$10$^{12}$ cm$^{−2}$ we extracted from the integrated rotational diagram analysis (see the second panel, first row, of Fig.~\ref{fig:rot_diag}). In particular, we applied a disk averaged rotational diagram, to fit both for the excitation temperature and for the molecular column density, as we previously outlined in Sect.~\ref{subsec:N_av}.

The excitation temperature we obtained is $(15 \pm 4)$\,K, which is low if we consider that \ce{c-C3H2} is expected to originate in the gas phase in the disk atmosphere  \citep{loison2017interstellar}, while $T_\mathrm{ex}\sim$15~K in the PDS~70 disk implies that the emitting layer is below $z/r=0.2$, under the assumption of LTE, and of a geometrically thin emitting layer. Nevertheless, the low excitation temperature could also result from the drastically different emitting radii of the two \ce{c-C3H2} transitions, while the rotational diagram assumes a uniform temperature. Another explanation could be that the \ce{c-C3H2} is actually originating from a higher layer, but the LTE approximation is not valid anymore and the resulting excitation temperature is lower.
On the other hand, both high and low emitting layers for the \ce{c-C3H2} molecule have been presented in other works: for example, \cite{cleeves2021tw} extracted $0.25<z/r<0.4$ for the TW Hya disk, while \cite{paneque2023directly} obtained $z/r\sim0.1$ for the HD~163296 disk. More observations and detailed chemical modeling are needed to shed light on the emitting layers of these molecules to constrain their formation pathways.

\subsection{No accretion signatures detected in SO}\label{subsec:faint_lines_SO}

The sulfur monoxide has been previously linked to a variety of processes occurring in planet forming disks \citep{sakai2014change, tabone2017alma}, including planetary accretion \citep{booth2022sulphur}. Nevertheless, the sulfur monoxide has not been detected in many class II disks, except for the disk around AB Aurigae \citep{pacheco2016high}, Oph-IRS~48 \citep{booth2021major}, HD~100546 \citep{booth2022sulphur},  and HD~169142 \citep{booth2023tracing,law2023so}. 

In the PDS~70 disk, we did not detect any signal from the SO molecule either localized onto the two actively accreting protoplanets or as an extended emission across the disk, after applying boosting techniques both in the $uv$ and in the image plane. We extracted an upper limit on the disk averaged column density of SO, from the $uv$-stacked data cube, obtaining $N_\mathrm{t}<2.7\times10^{12}$~cm$^{-2}$, assuming $T_\mathrm{ex}=30$\,K. Even if we rescale for the different disk masses (as $N_\mathrm{t} \propto$ disk mass) of the sources where SO has been detected and we consider the higher excitation temperatures they find (around 100\,K), the upper limit we extracted for the SO total column density is lower than the SO column density obtained for the sources listed above.

Additionally, we highlight the marginal detection of a localized 3$\sigma$ emission (white contour in the first panel of Fig.~\ref{fig:SO_signature}) in the channel map of the $uv$-stacked SO data cube, located at $\sim 1\farcs2$ outside the mm dust ring (pink contours), which is spatially and spectrally colocated with a kink in the $^{12}$CO Keplerian pattern (see the second panel) and with a gap in the H$_2$CO emission (see the third panel). Even if this detection is tentative, it would be interesting to further investigate a possible physical origin of this feature in high energy lines. Interestingly, the radial location of the observed SO marginal signature and of the H$_2$CO gap corresponds to the radial position of the peak intensity of CS and C$_2$H (see Fig.~\ref{fig:peak_profiles}), and to a dip in the brightness temperature profile of $^{12}$CO (\citealt{law2024mapping}). In this context, \cite{jiang2023chemical} showed that a giant planet forming inside a gap and outside the methane snowline can result in an enhanced gas-phase C/O ratio inside the gap.

\subsection{C/O ratio} \label{subsec:C/O}

The molecular line emission of the observed transitions is dominated by C-rich molecules, including various simple organic molecules, such as HCN, C$_2$H, CS, and \ce{c-C3H2}, while we did not detect CH$_3$OH and SO, which may suggest a gas phase O-depleted outer disk.
This evidence is fundamental in the context of two important astrochemical open questions related to the disk evolution, which are the understanding of the sulfur chemistry (main carriers, phase and evolution, \citealt{kama2019abundant}, \citealt{oberg2021astrochemistry}) and the fact that we observe a general depletion of volatile oxygen with respect to carbon in more evolved disks \citep{miotello2019bright, miotello2023setting}. 

In this context, there are several pieces of evidence pointing to an O-poor gas reservoir in the PDS~70 outer disk, and possibly to a higher C/O ratio, including:
\begin{itemize}
    \item bright emission of C$_2$H, which is enhanced in volatile O-depleted regions, preventing the destruction of the hydrocarbon \citep{bergin2016hydrocarbon, miotello2019bright, bosman2021molecules, alarcon2021molecules}, and the detection of \ce{c-C3H2}; 
    \item bright CS emission outside the submillimeter continuum ring, and the non-detection of SO, which has been previously linked to high C/O ratio \citep{guilloteau2016chemistry, semenov2018chemistry, le2021molecules, booth2021major, keyte2023azimuthal};
    \item non-detection of CH$_3$OH.
\end{itemize}

\cite{cridland2023planet} compared the observed radial profiles of \ce{^12CO}, \ce{C^18O}, and \ce{C2H} with the modeled ones obtained trough the thermo-chemical code DALI \citep{bruderer2012dali, bruderer2013dali}, in order to extract the volatile C/O ratio for the PDS~70 disk (we note that they inferred a star C/O ratio for PDS~70 of $\sim$0.44). The preferred model they obtained assumes a volatile C/O ratio marginally above unity, even if they suggest that radial variations of the C/O ratio across the disk may better reproduce both the inner peak and the outer emission shoulder in the observed radial profile of the \ce{C2H} integrated intensity. 

We can compare this result with the lower limit on the CS/SO ratio, which has been proposed to be a good direct tracer of the elemental C/O ratio \citep{guilloteau2016chemistry, semenov2018chemistry}. In particular, we extracted a lower limit on the CS/SO ratio, by taking the ratio of the CS average column density within a circle of radius 2$\as$ and the upper limit on the SO column density we obtained from the $uv$-stacked data cube. By varying the excitation temperature between 20-50\,K, we obtained a lower limit on the average CS/SO ratio between 0.8 and 1.1. \cite{semenov2018chemistry} explored the effect of the C/O ratio on the CS/SO one, through thermo-chemical models, obtaining that an increase in the volatile C/O ratio from 0.46 to 1.0 would result in raising the CS/SO ratio from $\sim0.026$ to $\sim13$. The inferred lower limit on the disk-averaged CS/SO ratio suggests that the gas phase C/O ratio in the outer disk may become greater than 1. With the cavity wall presenting a C/O ratio of 1 \citep{cridland2023planet}, this could be potentially indicating a radial gradient in the C/O ratio of the gas phase. In the future, a forward model extracting the gas C/O ratio from the very inner disk from the JWST data \citep{Perotti2023water} to the outer disk from the ALMA data presented in this paper should have high priority.

\section{Conclusions}\label{sec:conclusions}

In this work, we have presented high-resolution and deep-sensitivity ALMA band~6 line emission observations of the planet-hosting disk around PDS~70. In particular, we have presented an astrochemical overview of the molecular complexity of the PDS~70 disk, and searched for weak lines through image-plane and $uv$-plane boosting techniques. We summarize our conclusions below:

\begin{enumerate}
    \item We robustly detected ring-shaped \ce{^12CO}, \ce{^13CO}, \ce{C^18O}, \ce{H^13CN}, \ce{HC^15N}, DCN, \ce{H2CO}, \ce{C2H}, \ce{CS}, and \ce{H^13CO^+}, with the peak of the emission not always colocated with the millimeter continuum bright ring. Most of the molecules show a peak in the emission inside the millimeter dust ring, which we suggest is linked to the efficient irradiation of the cavity wall outside the planets' orbit on the disk chemistry; 
    \item We present the first detection of the \ce{c-C3H2} 218~GHz lines in the PDS~70 disk, leveraging boosting techniques;
    \item We suggest that the outer disk around PDS~70 has a O-depleted gas reservoir, supported by the non-detection of SO and CH$_3$OH, the detection of C-rich molecules (CS, C$_2$H, HCN, \ce{c-C3H2}), and a $\mathrm{CS/SO}\gtrsim1$ ratio;
    \item We extracted a radial profile of the excitation temperature and column density of \ce{H2CO}, from a radially resolved rotational diagram analysis. We propose a gas-phase formation pathway for the observed \ce{H2CO} emission, even if the grain-surface formation route cannot be completely ruled out;
    \item We have found evidence of an azimuthal asymmetry in the higher-energy \ce{H2CO} (3$_{2,1}$-2$_{2,0}$) line emission, which could be linked to accretion heating near PDS~70b. Moreover, we marginally detected a spatially and spectrally localized 3$\sigma$-signature in the SO emission, from the outer disk, colocated with a kink in the $^{12}$CO and a gap in the H$_2$CO peak intensity.
\end{enumerate}

Detailed chemical modeling is needed to quantify the C/O ratio across the dust cavity, shedding light on the composition of the gas accreted by the two giant planets, but also to test the suggested role of the irradiated cavity wall in shaping the molecular emission. 

\begin{acknowledgements}
We thank the referee for an exceptionally thorough review of the paper and very helpful comments. We are grateful to Ryan Loomis and Ian Czekala for the useful suggestions.
This paper makes use of the following ALMA data: ADS/JAO.ALMA\#2019.1.01619.S. ALMA is a partnership of ESO (representing its member states), NSF (USA) and NINS (Japan), together with NRC (Canada), MOST and ASIAA (Taiwan), and KASI (Republic of Korea), in cooperation with the Republic of Chile. The Joint ALMA Observatory is operated by ESO, AUI/NRAO and NAOJ.

L.R., S.F., and M.L. are funded by the European Union (ERC, UNVEIL, 101076613). Views and opinions expressed are however those of the authors only and do not necessarily reflect those of the European Union or the European Research Council. Neither the European Union nor the granting authority can be held responsible for them. S.F. also acknowledges financial contribution from PRIN-MUR 2022YP5ACE.

Support for C.J.L. was provided by NASA through the NASA Hubble Fellowship grant No. HST-HF2-51535.001-A awarded by the Space Telescope Science Institute, which is operated by the Association of Universities for Research in Astronomy, Inc., for NASA, under contract NAS5-26555.

The PI acknowledges the use of
computing resources from the Italian node of the European ALMA Regional
Center, hosted by INAF-Istituto di Radioastronomia.
\end{acknowledgements}

\bibliographystyle{aa}
\bibliography{references}

\onecolumn
\begin{appendix}

\section{Disk-integrated spectra}\label{appendix:int_spectra}
\begin{figure*}[h!]
    \includegraphics[scale=0.2, center]{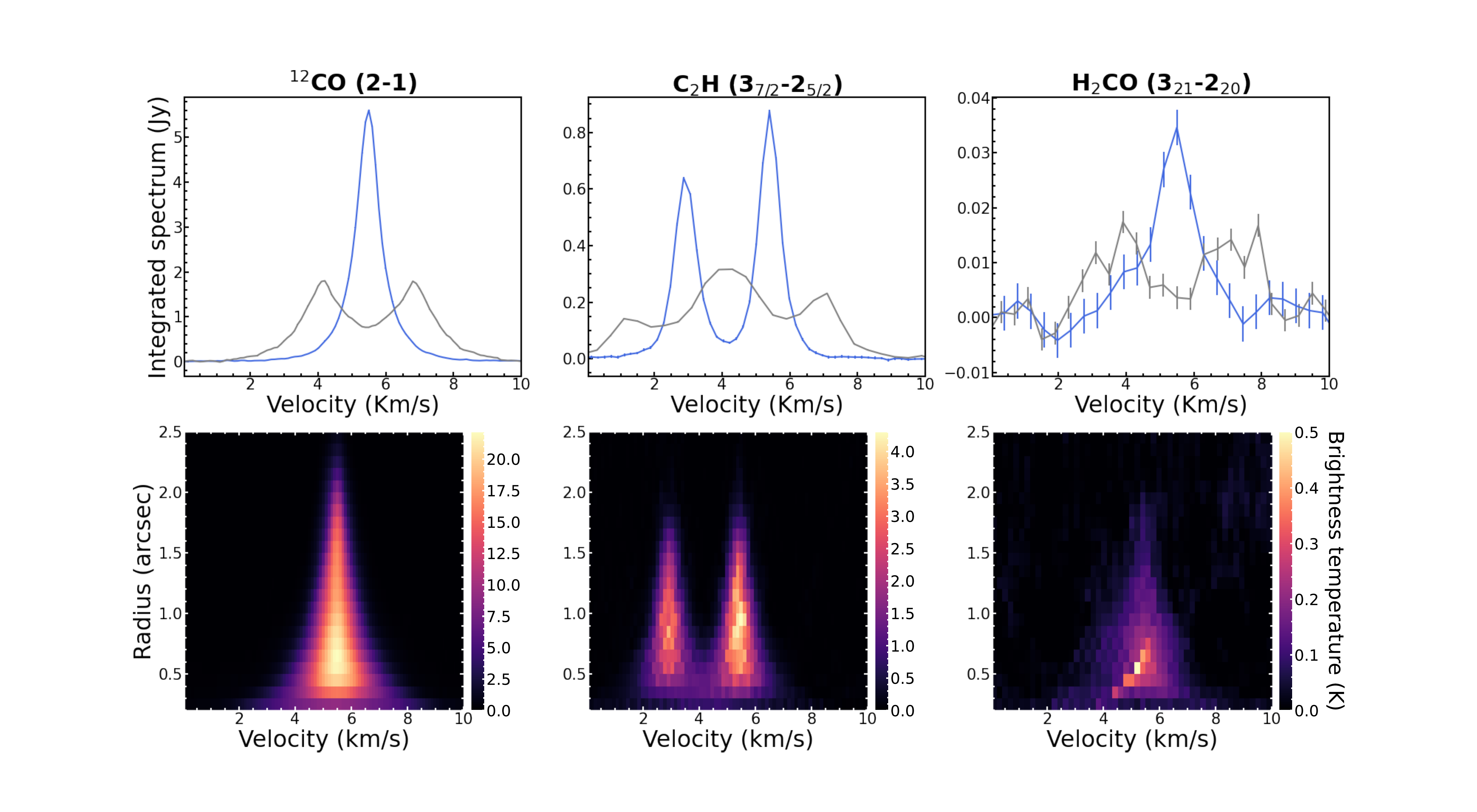}
     \caption{Effect of the Keplerian shifting on the line spectrum. Top row: Disk-integrated spectrum of $^{12}$CO ($J$=2-1), C$_2$H (3$_{7/2}$-2$_{5/2}$) and H$_2$CO ($3_{2,1}-2_{2,0}$), respectively, on a circular de-projected region of radius 3$\as$ for the $^{12}$CO and of 2$\as$ for the other ones. The blue line is the shifted spectrum, obtained by correcting for the Doppler shift due to the Keplerian rotation of the disk, while the gray line corresponds to the un-shifted spectrum.
     Bottom row: Spectra of $^{12}$CO ($J$=2-1), C$_2$H (3$_{7/2}$-2$_{5/2}$) and H$_2$CO (3$_{2,1}$-2$_{2,0}$) as a function of radius, obtained after the Keplerian shifting: the spectra are expressed in brightness temperature, using the Rayleigh-Jeans approximation. The hyperfine components of the C$_2$H transition are visible in the middle column.}
    \label{fig:spectra}
    \end{figure*}

\begin{figure*}[h!]
    \centering
    \includegraphics[scale=0.1]{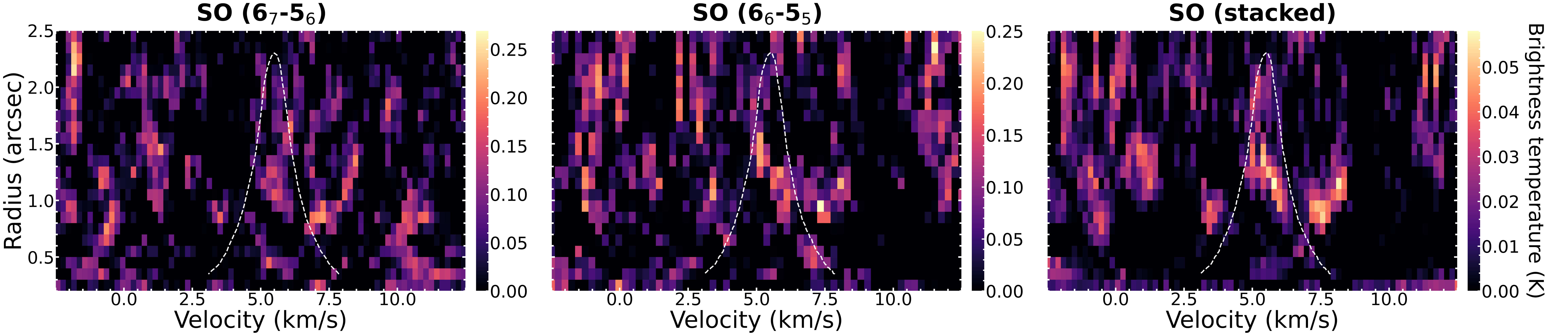}
    \caption{Spectra as a function of radius for the individual undetected lines SO 6$_7$-5$_6$ and 6$_6$-5$_5$, compared to the $uv$-stacked one. The white contours refer to the 3K level from the $^{12}$CO ($J$=2-1).}
    \label{fig:SO_uv_stacking}
\end{figure*}

We present the stacked spectra obtained leveraging the Keplerian rotation of the disk in Fig.~\ref{fig:spectra}, as we described in Sect.~\ref{subsec:fluxes} \citep{teague2016measuring}.
The top row in Fig.~\ref{fig:spectra} shows an example of three spectra for $^{12}$CO (2-1), C$_2$H (3$_{7/2}$-2$_{5/2}$) and H$_2$CO (3$_{2,1}$-2$_{2,0}$) transitions respectively: the shifted spectrum (in blue) visibly increases the S/N with respect to the original one (in gray), which shows the typical double horn effect due to the Doppler shift. The second column in Fig.~\ref{fig:spectra} shows the spectrally resolved hyperfine components of the C$_2$H transition, with the double horn of the hyperfine components being blended in the Doppler-shifted spectrum. We highlight the spatial distribution of the line emission by extracting the spatially integrated spectra as a function of radius, averaging over $0\farcs1$ wide annuli, after applying the Keplerian shifting: the result is presented in the bottom row of Fig.~\ref{fig:spectra}, for the $^{12}$CO (2-1), C$_2$H (3$_{7/2}$-2$_{5/2}$), and H$_2$CO ($3_{2,1}-2_{2,0}$) lines. Figure~\ref{fig:SO_uv_stacking} shows the Keplerian shifted spectra as a function of radius, for SO 6$_7$-5$_6$, 6$_6$-5$_5$, and for the $uv$-stacked cube. The white contours refer to the 3K level from the $^{12}$CO (2-1), which highlights the absence of a clear Keplerian emission.

\FloatBarrier

\section{Channel maps of the faint lines}

\begin{figure*}[h!]
    \centering
    \includegraphics[scale=1.2]{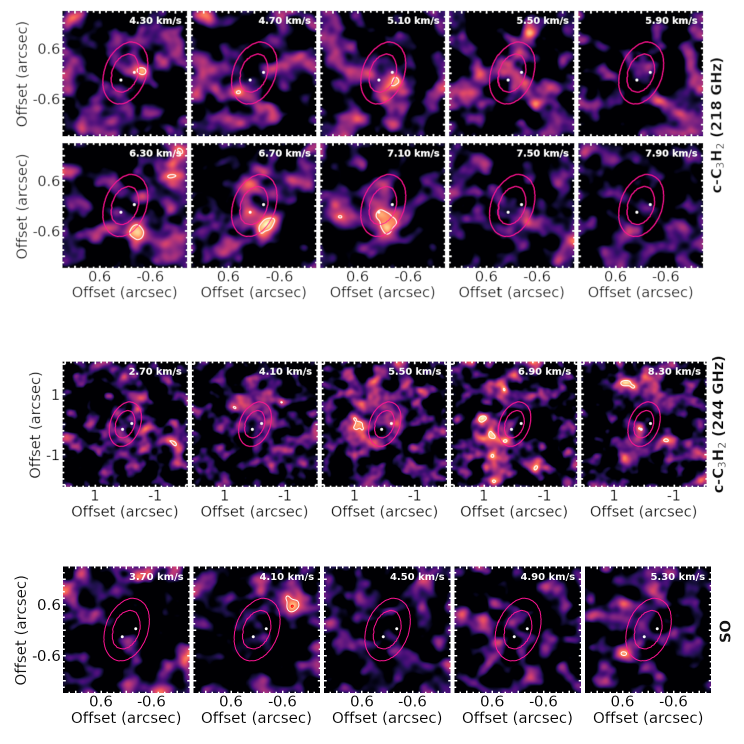}
    \caption{Channel maps of \ce{c-C3H2} (218 GHz, top row), c-C$_3$H$_2$ (244 GHz, middle row) and SO stacked lines (bottom row), with the white contours showing the 3$\sigma$ levels and the pink contours which indicate the extent of the bright ring in the band 7 continuum emission \citep{isella2019detection}. The white dots correspond to the position of the two protoplanets \citep{wang2021constraining}.}
    \label{fig:channel_maps}
\end{figure*}

In Fig.~\ref{fig:channel_maps} we present the channel maps of the detected \ce{c-C3H2} (218 GHz) and c-C$_3$H$_2$ (244 GHz) lines, and of the marginally detected SO, after the $uv$-stacking. The marginal SO signature in the outer disk, which we discussed in Sect.~\ref{subsec:faint_lines_SO}, is highlighted by the 3$\sigma$ white contour and by the 4$\sigma$ dot, at a velocity channel of 4.10~km~s$^{-1}$, in the last row. The channel maps of the faint \ce{c-C3H2} lines (first and middle rows) show 3$\sigma$ features at different velocity channels. The marginal detection of the signal from these faint lines throughout the channel maps highlights the importance of using boosting techniques, such as the ones we presented in this work, in order to asses the robustness of the detection.

\FloatBarrier

\section{Test on \texorpdfstring{$uv$}{uv}-stacking}

\begin{figure*}[h!]
    \centering
    \includegraphics[scale=0.55]{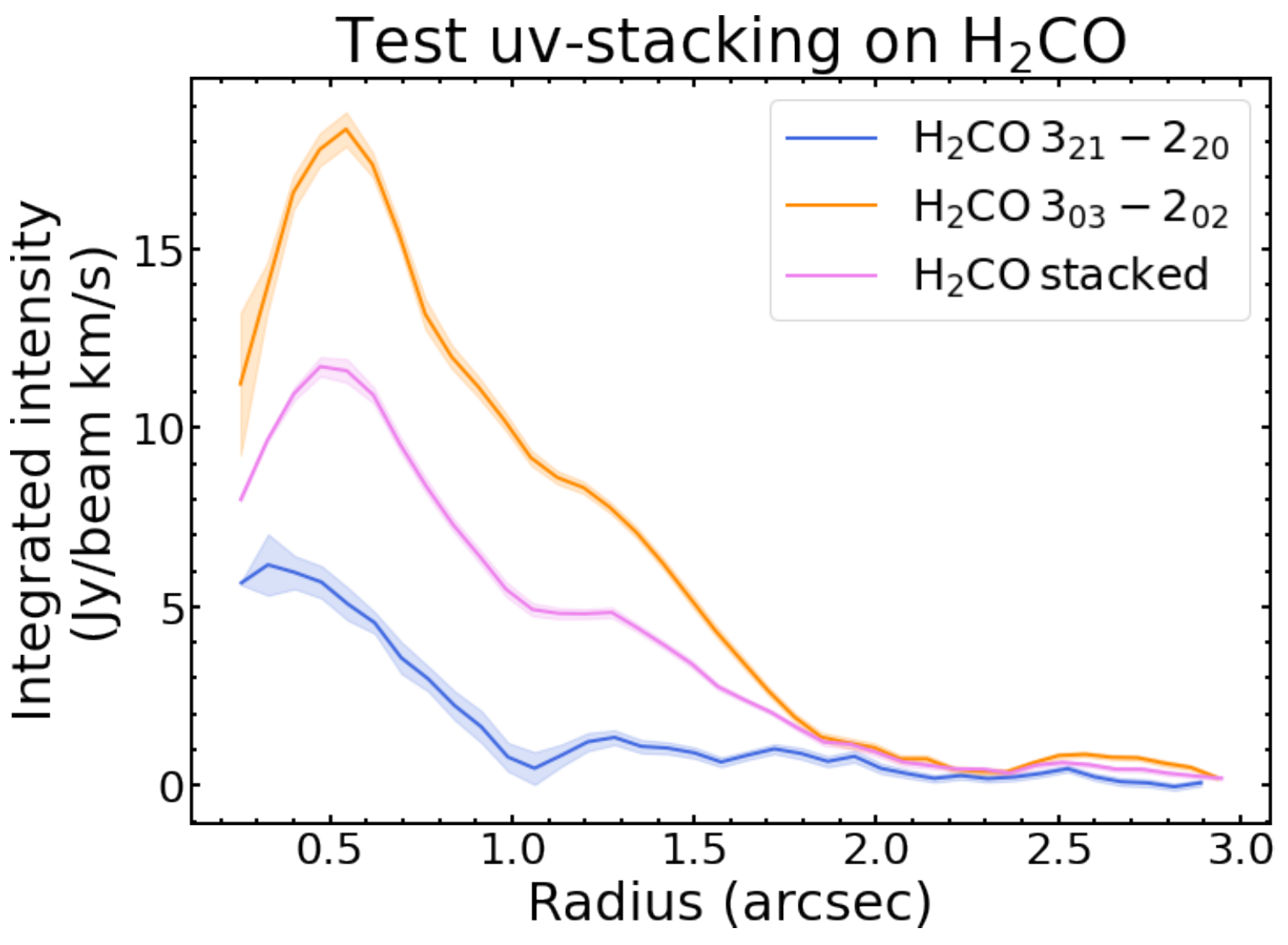}
    \caption{Test of the $uv$-stacking on the two individually detected lines of H$_2$CO, showing the radial profiles of the integrated intensity from the stacked cube (pink), which lies between the bright (orange) and faint (blue) ones, as expected.}
    \label{fig:test_uv_stacking}
\end{figure*}

We tested the effect of the $uv$-stacking, we presented in Sect.~\ref{subsec:stacking_SO}, on the two individually detected lines of H$_2$CO: the result is presented in Fig.~\ref{fig:test_uv_stacking}. As expected from the effect of the \texttt{CASA} task \texttt{concat} (see Sect.~\ref{subsec:stacking_SO} for more details), the integrated intensity profile of the stacked cube (pink line) lies between the higher and the lower profiles, representing the integrated intensity profiles of the bright (orange line) and the faint (blue line) transitions, respectively.
\FloatBarrier

\section{Gaussian fits to the radial profiles}

\begin{figure*}[h!]
    \centering
    \includegraphics[scale=0.20]{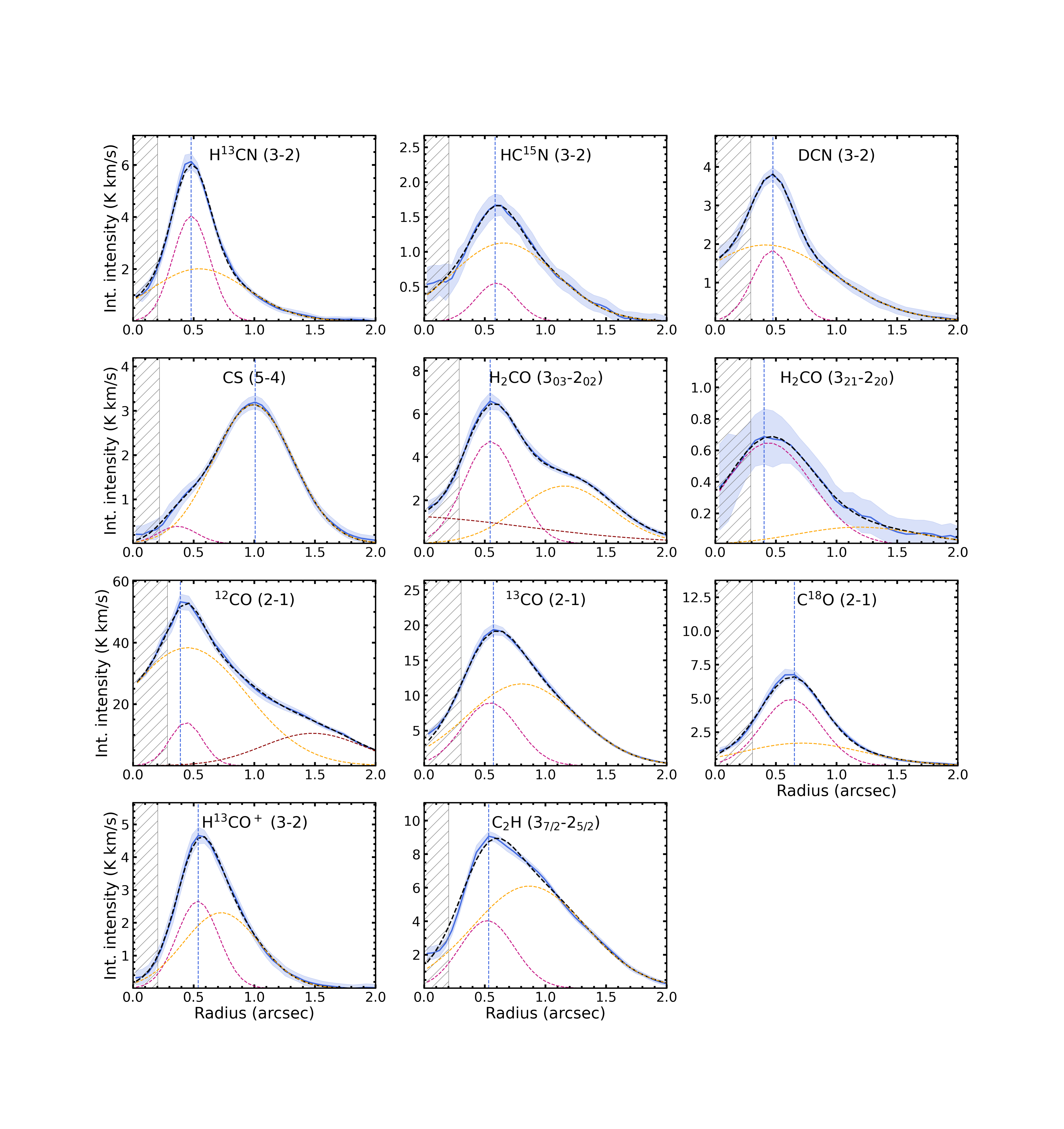}
    \caption{Radial profiles of the integrated intensity for the analyzed molecular transitions (blue lines), and corresponding multiple-Gaussian fit (dashed black line). Dashed orange, pink, and red lines show single Gaussian, which summed together result in the dashed black profiles. The vertical dashed blue line corresponds to the peak position directly inferred from the solid blue profile.}
    \label{fig:profiles_fit}
\end{figure*}

As we described in Sect.~\ref{subsec:discuss_lines}, we performed a multiple-Gaussian fit (triple for\ce{^12CO} and \ce{H2CO}, double for the other molecules) of the radial profiles of the molecular integrated intensity, in order to retrieve the radial location of the peak of the emission, and related uncertainty. We performed the fit procedure using \texttt{scipy.optimize.curve\_fit}, and the result is plotted in Fig.~\ref{fig:profiles_fit} (dashed black line), together with the single Gaussians (dashed orange, pink, and red lines). The fit is in very good agreement with the radial profiles extracted from observations (solid blue lines).

\begin{figure*}[h!]
    \includegraphics[scale=0.20]{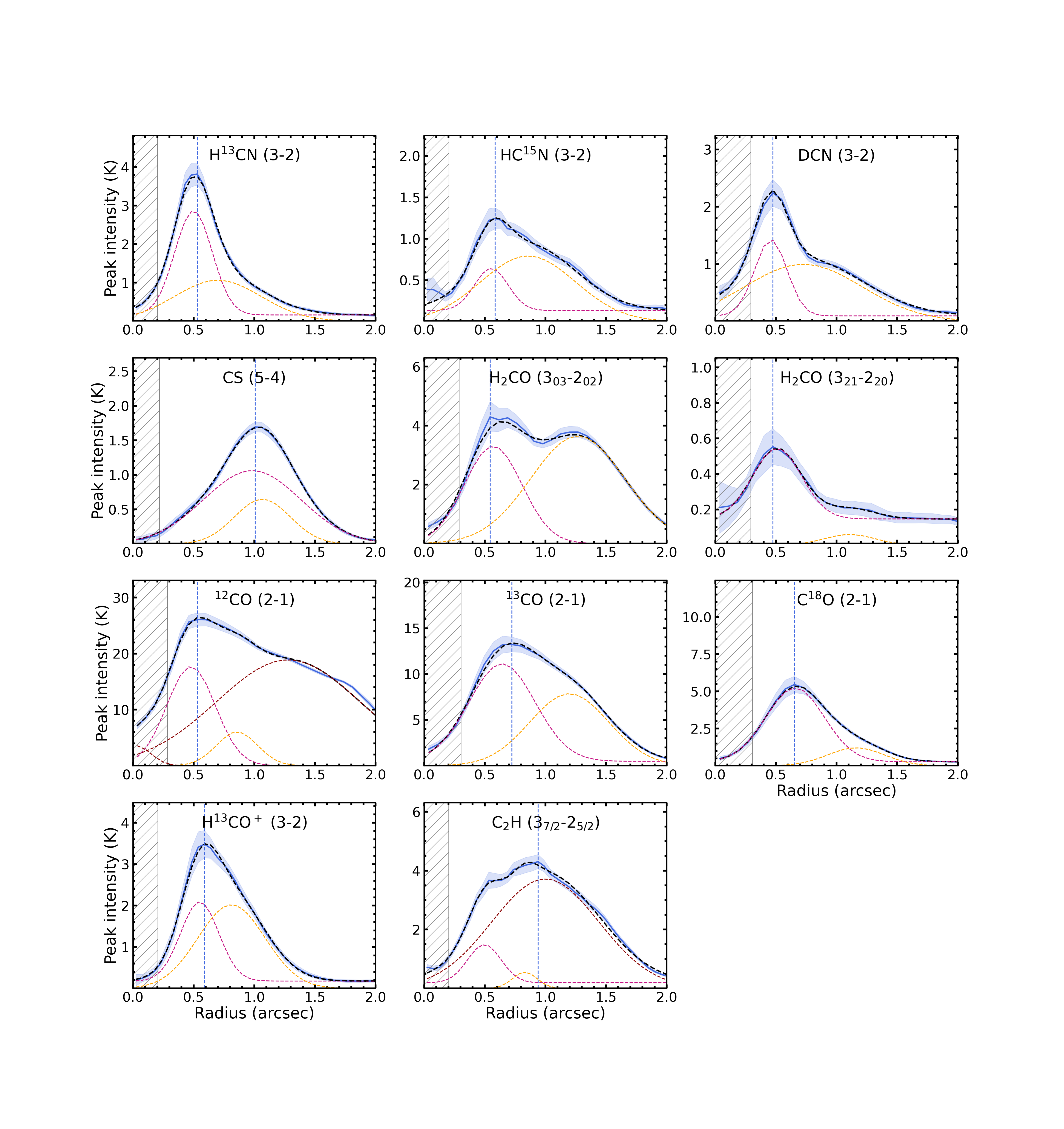}
    \caption{Same as Fig.~\ref{fig:profiles_fit} but for the peak intensity.}
    \label{fig:peak_profiles}
\end{figure*}

We also tested the multiple-Gaussian fitting to the peak intensity radial profiles (see Fig.~\ref{fig:peak_profiles}), and we listed the peak positions with associated uncertainties in Table~\ref{tab:radii_peaks}. In Fig.~\ref{fig:peak_cont_peak} we compare the peak positions extracted from the integrated intensity (marked in white) and from the peak intensity radial profiles (marked in red). There is good agreement between the two methods, except for the \ce{C2H} and \ce{^13CO} molecules.

	

\FloatBarrier

\begin{table}[h!]
\begin{minipage}[c]{0.45\linewidth}
\renewcommand{\arraystretch}{1.3}
	\centering
	\caption{
    Radial position of the peak of the molecular emission.}
	
	\label{tab:radii_peaks}
	\begin{threeparttable}
		\begin{tabular}{c c c }
			\hline
			\hline
			Transition\tnote{a} & $R_\mathrm{peak}$ \tnote{b} [$\as$] & $R^\mathrm{out}_\mathrm{peak}$ \tnote{c} [$\as$] \\
			\hline
			$^{12}$CO (2-1) & 0.48 $\pm$ 0.07 & - \\
			$^{13}$CO (2-1) & 0.64 $\pm$ 0.04 & - \\
			C$^{18}$O (2-1) & 0.66 $\pm$ 0.03 & - \\
			H$_2$CO (3$_{0,3}$-2$_{0,2}$) & 0.57 $\pm$ 0.02 & 1.27 $\pm$ 0.01 \\
			H$_2$CO (3$_{2,1}$-2$_{2,0}$) & 0.51 $\pm$ 0.03 & 1.11 $\pm$ 0.15 \\
			C$_2$H ($3_{7/2}$-$2_{5/2}$) & 0.83 $\pm$ 0.03 & - \\
			H$^{13}$CN (3-2) & 0.50 $\pm$ 0.01 & - \\
			HC$^{15}$N (3-2) & 0.55 $\pm$ 0.02 &- \\
			DCN (3-2) & 0.46 $\pm$ 0.01 & - \\
			H$^{13}$CO$^+$ (3-2) & 0.55 $\pm$ 0.01 & - \\
			CS (5-4) & 0.98 $\pm$ 0.02 & -\\
			\hline
			\hline
		\end{tabular}
		\begin{tablenotes}
			\item[a] See table \ref{tab:imaging_molecules} for quantum number formatting.
            \item[b] Peaks position were obtained from multiple-Gaussian fitting to the peak intensity radial profiles, for each molecule listed in the first column.
            \item[c] Radial position of possible second peaks.
		\end{tablenotes}
	\end{threeparttable}
\end{minipage}\hfill
\begin{minipage}[c]{0.54\linewidth}
	\centering
    \includegraphics[width=0.95\textwidth]{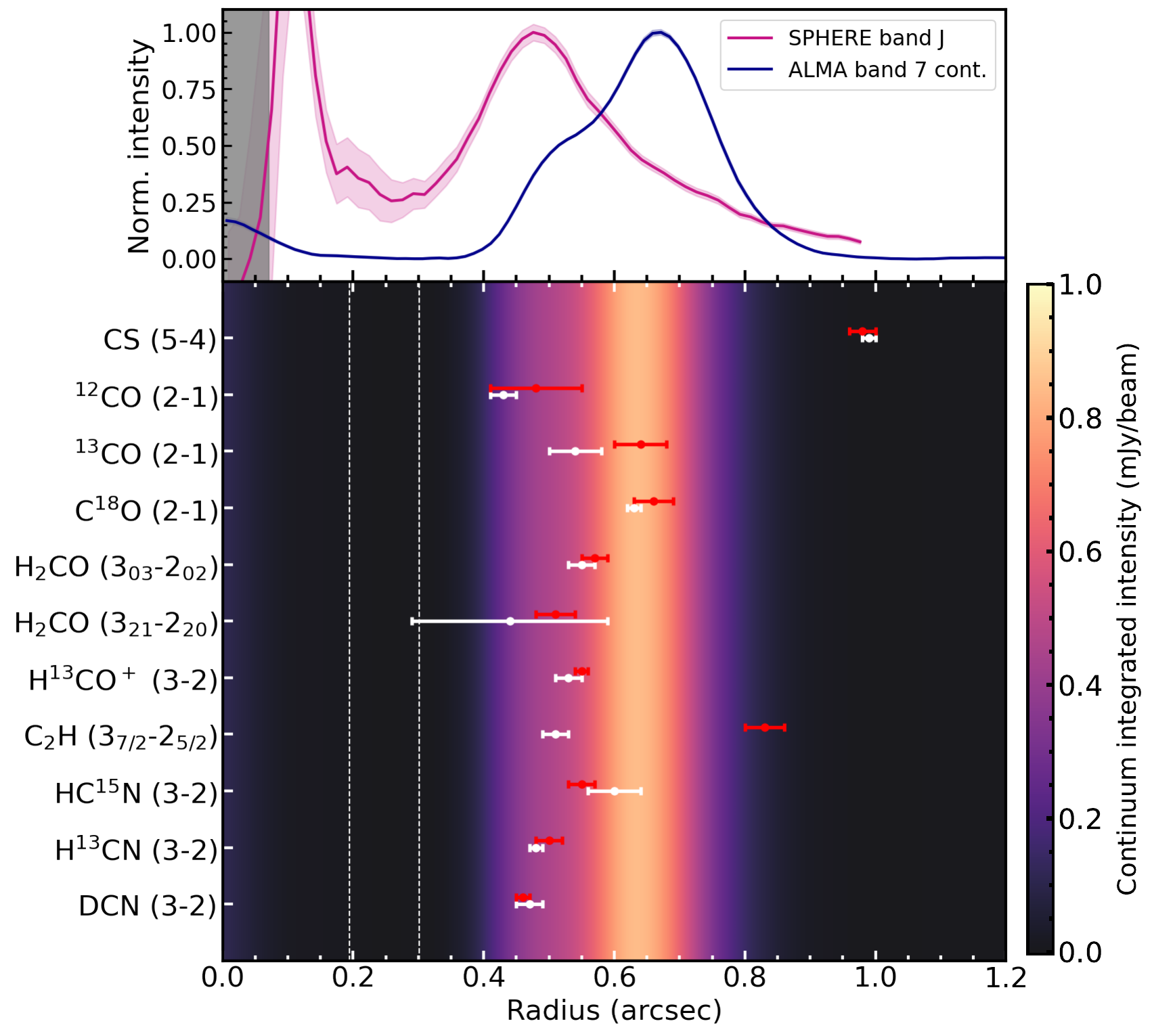}
	\captionof{figure}{Same as Fig.~\ref{fig:peak_cont}, with a comparison between the radial position of the peak of the emission extracted from the radial profiles of the integrated intensity (marked in white), and from the radial profiles of the peak intensity (marked in red). Peaks and associated uncertainties were obtained from multiple-Gaussian fitting (see Fig.~\ref{fig:profiles_fit} and \ref{fig:peak_profiles}).}
	\label{fig:peak_cont_peak}
\end{minipage}
\end{table}

\section{Peak intensity maps}

\begin{figure*}[h!]
    \centering
    \includegraphics[scale=0.20]{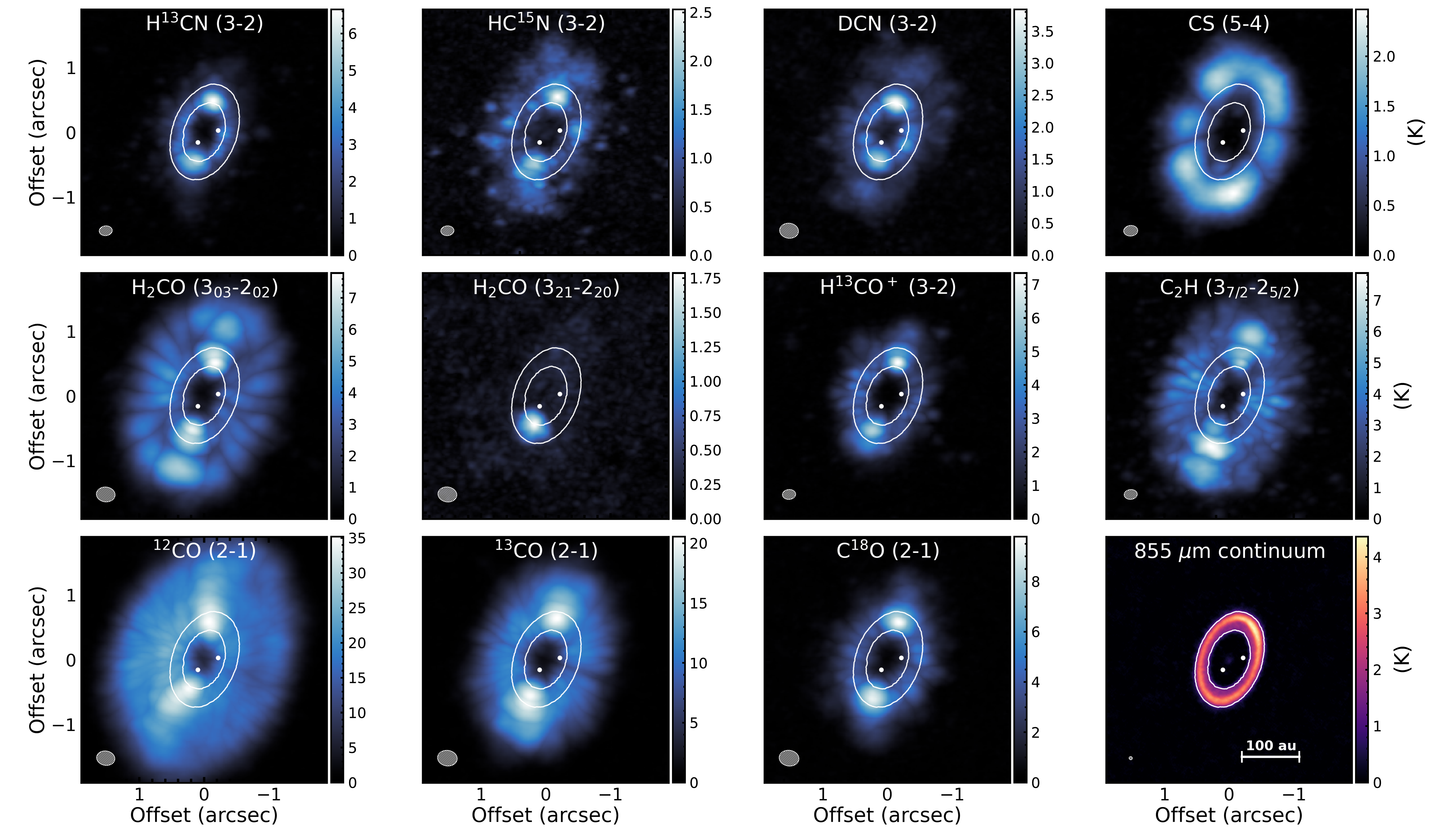}
    \caption{Peak intensity maps of the detected lines, with the white contours showing the bright ring in the submillimeter continuum emission \citep{isella2019detection} and the white dots marking the position of the two forming planets \citep{wang2021constraining}. The brightness temperature is derived in the Rayleigh-Jeans regime. The white ellipse on the bottom left of each panel is the beam.}
    \label{fig:M8}
\end{figure*}

We present the peak intensity maps of the same lines as in Fig.~\ref{fig:M0}, using the quadratic method implemented in the \texttt{bettermoments} tool \citep{teague2019bettermoments}. We extracted the maps from the JvM-corrected and continuum-subtracted cubes imaged with natural weighting. The result is presented in Fig.~\ref{fig:M8}. We present the result in K, assuming the Rayleigh-Jeans approximation.

The structured emission morphology is visible in the peak intensity maps, as in the integrated intensity maps, with the peak of the emission not always colocated with the submillimeter dust ring. Most of the molecules show axisymmetric emission, except for the 2.5$\sigma$ asymmetry along the major axis in H$_2$CO (3$_{2,1}$-2$_{2,0}$). The outer shoulders of emission of HC$^{15}$N, DCN, H$_2$CO and C$_2$H, outside the outer edge of the submillimeter ring, are more evident in the peak intensity maps, with respect to the integrated intensity maps, and they can be clearly seen in the corresponding radial profiles, in Fig.~\ref{fig:peak_profiles}. The optically thick $^{12}$CO emission shows an arced feature along the major axis in the peak intensity map, highlighting the elevated emitting layer.

The radial dark features are artifacts due to sparse channelization: for reference, these are particularly visible in the maps of CS and \ce{H2CO} (3$_{0,3}$-2$_{0,2}$). The azimuthal arced features could also be linked to physical and chemical processes affecting the molecular emission: these are particularly evident in the maps of \ce{C2H} and \ce{H2CO} (3$_{0,3}$-2$_{0,2}$), along the major axis, outside the bright continuum ring.

\FloatBarrier

\end{appendix}

\end{document}